\documentclass[journal]{IEEEtran}
\usepackage{amsmath,amsfonts}
\usepackage{algorithmic}
\usepackage{algorithm}
\usepackage{array}
\usepackage[caption=false,font=normalsize,labelfont=sf,textfont=sf]{subfig}
\usepackage{textcomp}
\usepackage{stfloats}
\usepackage{url}
\usepackage{verbatim}
\usepackage{graphicx}
\usepackage{cite}

\ifCLASSINFOpdf
\else
   \usepackage[dvips]{graphicx}
\fi
\usepackage{url}

\usepackage{amssymb}
\usepackage{latexsym}
\usepackage{color}
\usepackage{boldline}
\usepackage{multirow}

\newcommand{\F}{\ensuremath{\mathbb F}}
\newcommand{\E}{\ensuremath{\mathbb E}}
\newcommand{\Z}{\ensuremath{\mathbb Z}}
\newcommand{\C}{\ensuremath{\mathbb C}}
\newcommand{\R}{\ensuremath{\mathbb R}}

\newcommand{\mC}{\mathcal{C}}

\newcommand{\mI}{\mathcal{I}}

\newcommand{\mN}{\mathcal{N}}
\newcommand{\mO}{\mathcal{O}}

\newcommand{\mT}{\mathcal{T}}

\newcommand{\mV}{\mathcal{V}}
\newcommand{\mW}{\mathcal{W}}
\newcommand{\mX}{\mathcal{X}}

\newcommand{\mZ}{\mathcal{Z}}

\newcommand{\abu}{{\bf a}}

\newcommand{\cbu}{{\bf c}}

\newcommand{\ebu}{{\bf e}}

\newcommand{\hbu}{{\bf h}}

\newcommand{\sbu}{{\bf s}}

\newcommand{\vbu}{{\bf v}}

\newcommand{\xbu}{{\bf x}}
\newcommand{\ybu}{{\bf y}}
\newcommand{\wbu}{{\bf w}}

\newcommand{\Hbu}{{\bf H}}
\newcommand{\Fbu}{{\bf F}}
\newcommand{\Ibu}{{\bf I}}

\newcommand{\Sbu}{{\bf S}}
\newcommand{\Gbu}{{\bf G}}

\newcommand{\Wbu}{{\bf W}}
\newcommand{\Xbu}{{\bf X}}
\newcommand{\Ybu}{{\bf Y}}
\newcommand{\Phibu}{{\bf \Phi}}
\newcommand{\bGamma}{{\bf \Gamma}}
\newcommand{\bgam}{{\boldsymbol \gamma}}
\newcommand{\phibu}{{\boldsymbol \phi}}

\newcommand{\bSigma}{{\boldsymbol \Sigma}}

\newcommand{\iproof}{{\noindent \textit{Proof}}}

\newcommand{\qed}{\hfill \ensuremath{\Box}}

\newtheorem{fact}{Fact}

\newtheorem{df}{Definition}
\newtheorem{thr}{Theorem}
\newtheorem{lem}{Lemma}
\newtheorem{rem}{Remark}

\newtheorem{const}{Construction}

\begin{document}

\title{Joint Activity and Data Detection for Massive Grant-Free Access Using Deterministic Non-Orthogonal Signatures}

\author{Nam Yul Yu,~\IEEEmembership{Senior Member,~IEEE} and Wei Yu,~\IEEEmembership{Fellow, IEEE}
\thanks{
The work of Nam Yul Yu was supported by
the National Research Foundation of Korea (NRF) grant funded by the Korea Government (MSIT) (NRF-2022R1F1A1066143).
The work of Wei Yu was supported by
the Natural Sciences and Engineering Research Council (NSERC) of Canada via a Discovery Grant.}
\thanks{Nam Yul Yu is with the School of
of Electrical Engineering and Computer Science (EECS), Gwangju Institute of Science and Technology (GIST), Gwangju, 61005, Korea
(e-mail: nyyu@gist.ac.kr).}
\thanks{Wei Yu is with the Edward S. Rogers Sr. Department of Electrical and Computer Engineering,
University of Toronto, Toronto, ON, M5S 3G4, Canada
(email: weiyu@comm.utoronto.ca).}
}  

\markboth{Journal of \LaTeX\ Class Files,~Vol.~14, No.~8, August~2021}%
{Shell \MakeLowercase{\textit{et al.}}: A Sample Article Using IEEEtran.cls for IEEE Journals}


\maketitle

\begin{abstract}
Grant-free access is a key enabler for connecting 
wireless devices with low latency and low signaling overhead in 
massive machine-type communications (mMTC). 
For massive grant-free access,
user-specific signatures are uniquely assigned to mMTC devices.
In this paper,
we first derive a sufficient condition 
for the successful identification of active devices 
through maximum likelihood (ML) estimation
in massive grant-free access.
The condition is represented by the coherence 
of a signature sequence matrix containing the signatures of all devices.
Then, we present a design framework of non-orthogonal signature sequences in a deterministic fashion.
The design principle relies on unimodular masking sequences with low correlation, 
which are applied as masking sequences to the columns of the discrete Fourier transform (DFT) matrix.
For example constructions, 
we use four polyphase masking sequences represented by characters over finite fields. 
Leveraging algebraic techniques,
we show that the signature sequence matrix of proposed non-orthogonal sequences 
has theoretically bounded low coherence. 
Simulation results 
demonstrate that the deterministic non-orthogonal signatures achieve the excellent performance 
of joint activity and data detection by ML- and approximate message passing (AMP)-based algorithms
for massive grant-free access in mMTC.
\end{abstract}

\begin{IEEEkeywords}
Characters, coherence, 
grant-free access, massive machine-type communications, non-orthogonal signatures.
\end{IEEEkeywords}

\IEEEpeerreviewmaketitle

\section{Introduction}

\IEEEPARstart{M}{assive} 
machine-type communications (mMTC) is an important use case
of 5G and beyond wireless technology for
concretizing the Internet of Things (IoT)~\cite{Bockel:mMTC}. 
In mMTC,
only a small fraction of the massive number of wireless devices 
attempt to access a base station (BS) with no access grant.
The \emph{grant-free} access enables massive connectivity
with low latency and low control overhead~\cite{Cirik:toward}. 
For uplink grant-free access,
signature sequences are uniquely assigned to mMTC devices in a cell
so that each active device sends its own signature as a pilot 
in an access trial.
Then, a BS receiver tries to identify active devices,
to estimate channel profiles, and/or to detect transmitted data 
from the superimposed signatures~\cite{Liu:massive, Liu:mimo, Liu:mimo2, Jiang:noma, Cui:mimo, Larsson:grant}.

In mMTC, uplink grant-free access
can be accomplished by a two-phase access scheme\cite{Liu:massive, Liu:mimo, Liu:mimo2, Cui:mimo}.
In the first phase,
the BS receiver identifies active devices and estimates their channel profiles 
jointly from the superimposed signatures. 
Then in the second phase, it detects each active device's data 
followed by the signature, exploiting the estimated channel profile.
Since signature sequences
are uniquely assigned to mMTC devices, 
the number of non-orthogonal signatures required for
this two-phase scheme should be at least
the total number of devices in a cell.
In addition, 
a non-coherent access scheme~\cite{Larsson:grant} 
has been proposed by 
embedding data symbols in signature sequences.
The non-coherent scheme can be more efficient for massive access, 
since identifying signatures at a BS allows
joint activity and data detection in a single phase,
with no need of channel acquisition.
However, this scheme has to uniquely assign
a set of multiple signature sequences to each device,
which requires the number of non-orthogonal signatures to be larger than
the number of devices.

When signature sequences are uniquely assigned to devices,
the problem of activity detection at a BS 
boils down to identifying signature sequences transmitted from active devices.
In many research works, the \emph{sequence identification problem}
has been tackled by the technique of compressed sensing (CS)~\cite{Eldar:CS}. 
A variety of algorithms and methods have been deployed
in~\cite{Wang:struct, Wei:amp, Du:efficient, Du:joint, Cirik:admm, Du:block, Liu:massive, Liu:mimo, Liu:mimo2, Larsson:grant, 
	Jiang:noma, Shao:dim, Ke:meet, Mei:CS, Hara:hyper, Jiang:freq, Wei:multi, Chen:dct} to solve the problem
under a CS framework.
For example,
several algorithms based on greedy pursuit~\cite{Tropp:somp, Dai:sup}, 
approximate message passing (AMP)~\cite{Donoho:amp, Ye:BP, Schniter:mmv}, 
and Bayesian learning~\cite{Wipf:M-SBL} 
have been used as detection techniques for the problem. 
Recently, the maximum likelihood (ML) estimation
has been formulated to solve the problem 
using a sample covariance matrix of received signals~\cite{Caire:Non}.
Algorithms based on the coordinate descent method~\cite{Yu:cov, Cui:stat}
show excellent performance for solving the problem of ML estimation.
Moreover, the asymptotic behavior of ML estimation
has been studied by numerical analysis of the phase transition in~\cite{Yu:phase}, where
a necessary and sufficient condition is presented
for an accurate solution of the ML estimation employing 
an asymptotically large number of BS antennas.

In massive grant-free access,
a variety of user-specific, non-orthogonal sequences 
have been considered for signatures, pilots, or spreading, where the number of sequences
is much more than the sequence length. 
In literature,
various kinds of randomly generated non-orthogonal sequences, 
e.g.,
random complex Gaussian\cite{Liu:mimo, Jiang:noma, Shao:dim, Cui:mimo, Yu:cov, Cui:stat, Yu:phase, Ke:meet, Wei:multi, Chen:dct},
random QPSK~\cite{Larsson:grant}, 
unimodular with random phase~\cite{Hara:hyper}, 
and random partial Fourier~\cite{Jiang:freq, Mei:CS} sequences
have been utilized. 
Also, complex-valued non-orthogonal sequences randomly taking
finite elements have been used for multi-user shared access (MUSA)~\cite{Yuan:MUSA}.
Besides random ones, several \emph{deterministic} sequences have been proposed
for non-orthogonal multiple access (NOMA),
including
pseudo-random noise~\cite{Wang:struct, Wei:amp, Du:efficient, Du:joint, Cirik:admm, Du:block},
sinusoidal~\cite{Hasan:sinu}, and
Golay based~\cite{Yu:binary, Yu:non} sequences.
The Zadoff-Chu (ZC) sequences~\cite{Chu:ZC}
have been adopted as preambles for random access in 3GPP-LTE~\cite{3gpp:36.211}.
Generated in a systematic and structured way,
deterministic sequences enable 
efficient implementation in practice.

Exploiting the results of~\cite{Yu:phase}, 
this paper 
first derives a sufficient condition for the ML estimation 
to achieve a true solution of the sequence identification problem.
While the asymptotic analysis of~\cite{Yu:phase} relies on numerical experiments of the phase transition,
the sufficient condition of this paper is simply represented by 
the \emph{coherence} of a signature sequence matrix 
that contains the non-orthogonal signatures of length $L$ from all devices.
In particular, the condition is useful
when the non-orthogonal signature sequences are
generated in a deterministic manner.
The sufficient condition suggests that 
the ML estimation can identify at least $K = \mO(L)$ transmitted signatures successfully with high probability
using an asymptotically large number of BS antennas.
The condition highlights the importance of low coherence of the signature sequence matrix 
for reliable sequence identification.

From our coherence-based analysis,
we need a set of non-orthogonal signatures
for which the coherence of the corresponding signature sequence matrix 
is as low as possible. 
Also, 
it is important to design a large number of signature sequences
for the non-coherent access scheme~\cite{Larsson:grant}
to accommodate a massive number of devices. 
For this purpose,
we first present a design framework for deterministic signature sequences of length $L$,
where unimodular masking sequences are applied in a masking operation  
to the columns of the $L$-point discrete Fourier transform (DFT) matrix.
For example constructions, we utilize four polyphase masking sequences,
where the elements are represented by characters~\cite{Lidl:FF}
over finite fields.
In specific, we consider cubic~\cite{Alltop:comp, Ye:LAZ}
and trace~\cite{Sidel:mutual, HellKumar:low} sequences,
represented by additive characters.
Also, we use power residue and Sidelnikov sequences~\cite{Sidel:org},
represented by multiplicative characters.

By applying the above masking sequences,
the deterministic design is able to supply 
$ \mO(L^3)$ 
non-orthogonal sequences of length $L$.
Leveraging the bounds on character sums~\cite{Wang:new},
we show that the signature sequence matrix 
containing the non-orthogonal sequences
has the theoretically bounded low coherence of $\mO (\frac{1}{\sqrt{L}})$, 
nearly meeting the Welch bound equality~\cite{Welch:low}.
Our coherence-based analysis suggests that
the deterministic design, 
thanks to the low coherence, can guarantee reliable
ML estimation for the sequence identification problem with 
$K= \mO(L)$ transmitted signatures.
Moreover, simulation results reveal that
the ML estimation can identify the deterministic signatures accurately
even for $K>L$,
using a massive number of BS antennas.
Compared to randomly generated and algorithmically optimized ones, 
the proposed non-orthogonal signatures also enjoy low implementation cost, 
as generated efficiently in practice
by the deterministic design.

In simulations, we investigate the performance
of proposed non-orthogonal signatures in the non-coherent access scheme.
For joint activity and data detection,
we deploy the coordinate descent algorithm~\cite{Yu:cov} for
maximum likelihood estimation (CD-ML) and 
the approximate message passing (AMP) with multiple measurement vectors (MMV-AMP)~\cite{Ke:mmv-amp}, 
respectively.
Simulation results demonstrate that the proposed non-orthogonal signatures
achieve excellent performance for joint activity and data detection
in the non-coherent access scheme, 
outperforming several randomly generated signatures, e.g.,
random complex Gaussian, random MUSA, and random QPSK sequences.
In particular, the proposed non-orthogonal signatures of short lengths
have outstanding performance for joint activity and data detection using CD-ML
with a massive number of BS antennas,
which allows massive access 
with low signaling overhead.
In conclusion,
the proposed non-orthogonal signatures
are promising for massive grant-free access in mMTC,
thanks to the theoretically bounded low coherence
and the low implementation cost.

The main contributions of this paper are summarized as follows.
\begin{itemize}
	\item We derive a sufficient condition to achieve a true solution of the sequence identification problem
	through ML estimation.
	Given an arbitrary signature sequence matrix,
	we use the coherence to represent the condition.
	The sufficient condition gives a guideline on how 
	non-orthogonal 
	signatures should be designed to provide low coherence for the signature sequence matrix,
	which allows reliable ML estimation for the sequence identification problem.
	\item We construct four sets of deterministic non-orthogonal signature sequences 
	for massive grant-free access.
	To be used for the non-coherent access scheme,
	each sequence set provides a large number of sequences as well as low coherence
	for the signature sequence matrix.
	The design principle is based on the application
	of unimodular sequences for masking the columns of the DFT matrix.
	This deterministic design provides us with $\mO(L^3)$ non-orthogonal signature sequences of length $L$,
	where the coherence of the corresponding signature sequence matrix is theoretically bounded by $\mO(\frac{1}{\sqrt{L}})$.
	\item Simulation results demonstrate that the proposed non-orthogonal
	signatures show outstanding performance for joint activity and data detection
	in massive grant-free access.
	In particular, 
	the proposed signatures of short lengths
	support massive grant-free access with low signaling overhead,
	when the BS has a large number of antennas.
	Thanks to the low coherence and the low implementation cost,
	the proposed non-orthogonal signatures based on deterministic design
	can be suitable for massive grant-free access in mMTC.
\end{itemize}

The rest of this paper is organized as follows.
Section II outlines
a system model of the non-coherent access scheme,
where the problem of joint activity and data detection is formulated.
Also, we review the covariance-based ML estimation 
to solve the problem.
In Section III, we derive a sufficient condition 
to achieve a true solution of the sequence identification problem through ML estimation. 
Section IV presents a design framework 
of deterministic non-orthogonal signature sequences. 
With the design framework, we construct four sets of non-orthogonal signature sequences
by using cubic, power residue, Sidelnikov, and trace sequences for masking operation.
Section V presents simulation results to 
demonstrate the performance of proposed non-orthogonal signatures
for joint activity and data detection in massive grant-free access.
Finally, concluding remarks will be given in Section VI.

\noindent \textit{Notations}: 
In this paper,
a matrix (or a vector) is represented by a bold-face upper (or a lower) case letter.
The transpose and the conjugate transpose of a matrix $\Xbu$ are denoted by
$\Xbu^T$ and $\Xbu^H$, respectively.
Note that $\Xbu^*$ denotes a matrix having the conjugate elements of $\Xbu$
with no transpose.
The identity matrix is denoted by $\Ibu$,
where the dimension is determined in the context.
${\rm diag} (\hbu)$ is a diagonal matrix whose diagonal entries are from a vector $\hbu$.
The inner product of vectors $\xbu$ and $\ybu$ is denoted by $\langle \xbu, \ybu \rangle$.
The $l_2$-norm of a vector $\xbu  = (x_1, \cdots, x_N)$ is denoted by
$ || \xbu ||_{2} = \sqrt{ \sum_{k=1} ^{N} |x_k|^2 } $.
The Frobenius norm of a matrix $\Xbu $ is 
$\| \Xbu \|_F = \sqrt{\sum_{k, l} \left| \Xbu(k,l) \right| ^2 }$.
The coherence of a matrix $\Xbu$ is defined by 
$\mu(\Xbu) = \max_{ k \neq l  }
\frac{\left| \left \langle \xbu_{k} ,  \xbu_{l}  \right \rangle \right| }
{\| \xbu_{k} \|_2 \| \xbu_{l} \|_2}$, 
where $\xbu_k$ and $\xbu_l$ are the $k$th and the $l$th columns of $\Xbu$, respectively.
For a pair of vectors $\xbu$ and $\ybu$,
$\xbu \odot \ybu$ denotes the element-wise multiplication of $\xbu$ and $\ybu$.  
If $\xbu$ and $\ybu$ are binary vectors of elements $0$ and $1$, 
$\xbu \oplus \ybu$ denotes the bitwise XOR addition of $\xbu$ and $\ybu$. 
Finally, $\hbu \sim \mathcal{CN} (\mathbf{m}, \mathbf{\Sigma})$ is a circularly symmetric complex Gaussian random vector
with mean $\bf m$ and covariance $\mathbf{\Sigma}$.

\section{System Model}

\subsection{Grant-Free Random Access} 
In this paper, we consider uplink grant-free random access in single-cell mMTC, 
where a base station (BS) equipped with $M$ antennas accommodates
total $N_d$ single-antenna devices.
In each random access interval, 
only $K$ devices out of total actively transmit their own signatures of length $L$
to the BS synchronously~\cite{Liu:mimo}, where $K \ll N_d$ due to sparse activity.
In mMTC, small $L = \mO(K)$ would be desirable 
for massive grant-free access with low signaling overhead.
For fully grant-free access, 
we assume that devices are \emph{static}
in a cell and thus BS accommodates a fixed set of devices
having their own user-specific signatures\footnote{
IoT technologies such as low power wide area networks (LPWAN)~\cite{Raza:LPWAN}
typically assume that devices are static.
For example, 
small amount of data are sent by low-cost devices with no mobility management 
in the narrow-band IoT (NB-IoT)~\cite{Ratasuk:nbiot}, where  
the IoT devices are assumed to be static or immobile.}.

For efficient data transmission in grant-free random access, we consider
a non-coherent access scheme\cite{Larsson:grant, Yu:cov}
by embedding data information in signature sequences.
For this purpose,
a unique signature set $\Sbu_n \in \C ^{L \times Q}$ containing $Q$ distinct sequences
is allocated to
device $n$, i.e.,
\begin{equation}\label{eq:Sbu_n}
\Sbu_n = \left [ \sbu_n ^{(1)}, \cdots, \sbu_n ^{(Q)} \right ],
\end{equation}
where $\sbu_n ^{(q)} = (s_{n,1} ^{(q)}, \cdots, s_{n, L} ^{(q)})^T$ 
is a signature sequence of length $L$
for $ q = 1, \cdots, Q$ and $n=1, \cdots, N_d$.
Note that $L \ll N_d$ due to a massive number of devices in mMTC.
When device $n$ is active and wishes to send $J$ bits of information,
it transmits a single sequence out of $\Sbu_n$, where $Q = 2^J$.
Then, 
one can detect the activity and data of device $n$ simultaneously
by identifying the transmitted signature from $\Sbu_n$.

For device $n$,
an indicator vector can be defined by 
$\abu_n = (a_n ^{(1)}, \cdots, a_n ^{(Q)})^T $,
where $a_n^{(q)} \in \{ 0, 1\}$ indicates whether or not a sequence $\sbu_n ^{(q)}$ is transmitted.
Since only one sequence of $\Sbu_n$ is transmitted by active device $n$,
$\| \abu_n \|_1 = 1$ if device $n$ is active, 
and $ \| \abu_n \|_1 = 0$ otherwise.
Then, $\mI = \{ n \mid \|\abu_n \|_1 = 1, \ n = 1, \cdots, N_d \}$ is
a set of active devices, where
the number of active devices is $|\mI| = \sum_{n=1} ^{N_d} \| \abu_n \|_1= K \ll N_d$.
Note that we can detect the activity of device $n$  
by checking whether $\| \abu_n \|_1 = 0$ or $1$,
while data bits from active device $n$ are detected 
by the index $q$ of $a_n ^{(q)}=1$ from $\abu_n$.

Let $g_n \hbu_n  \in \C ^{M \times 1}$ 
be a channel vector for device $n$,
where $g_n$ is the large-scale fading component determined by the distance between device $n$ and the BS,
and
$\hbu_n \sim \mathcal{CN}(\bf{0}, \Ibu) $ is the Rayleigh fading gains over $M$ antennas.
Then, the received signal at the BS
can be represented by
\begin{equation}\label{eq:Y}
\begin{split}
\Ybu & = \sum_{n=1} ^{N_d} \sum_{q=1} ^Q a_n ^{(q)} g_n  \sbu_n ^{(q)} \hbu_n ^T + \Wbu \\
& = \sum_{n=1} ^{N_d} \Sbu_n {\rm diag} (g_n \abu_n) \Hbu_n   + \Wbu, 
\end{split}
\end{equation}
where $\Hbu_n = [\hbu_n, \cdots, \hbu_n]^T \in \C ^{Q\times M}$
is a channel matrix for device $n$ with $Q$ repeated rows of $\hbu_n ^T$.
In~\eqref{eq:Y}, 
$\Wbu \sim \mathcal{CN}({\bf 0}, \sigma_w ^2 \Ibu) $ is the complex Gaussian noise
with variance $\sigma_w ^2$.

Let $\Sbu = [\Sbu_1, \cdots, \Sbu_{N_d}] \in \C ^{L \times N}$
be a concatenation of the matrices~\eqref{eq:Sbu_n} across $N_d$ devices,
where $N = N_dQ$.
In~\eqref{eq:Y}, $\Ybu \in \C^{L \times M}$ can then be expressed in matrix form, i.e.,
\begin{equation}\label{eq:Y_mat}
\Ybu = \Sbu \bGamma ^{\frac{1}{2}} \Hbu + \Wbu,
\end{equation}
where $\bGamma ^{\frac{1}{2}} = {\rm diag} \left([g_1 \abu_1 ^T, \cdots, g_{N_d} \abu_{N_d} ^T] \right) \in \C^{N \times N}$
and
$\Hbu = [\Hbu_1^T \cdots, \Hbu_{N_d} ^T]^T \in \C^{N \times M}$.
In~\eqref{eq:Y_mat}, the diagonal vector of $\bGamma$ is denoted by
$\bgam = [\bgam_1 ^T, \cdots, \bgam_{N_d} ^T]^T  \in \C ^{N}$, where
$\bgam_n =  g_n^2 \abu_n $.
Finally, the BS receiver tackles the problem \eqref{eq:Y_mat}
to find $\bgam$, or $\bgam_n =  g_n^2 \abu_n $ for all $n = 1, \cdots, N_d$,
which ultimately detects
device activity and data simultaneously.


\subsection{Covariance-Based Maximum-Likelihood (ML) Estimation}
In this paper, we mainly consider 
the covariance-based maximum-likelihood (ML) estimation
to find $\bgam$ in~\eqref{eq:Y_mat}.

%

As the channel coefficients are i.i.d.,
each column of $\Ybu$ in~\eqref{eq:Y_mat},
denoted by $\ybu_m \in \C^L, 1 \le m \le M$, 
follows the complex Gaussian distribution independently,
i.e., $\ybu_m \sim \mathcal{ CN} ({\bf 0}, \bSigma)$,
where the covariance matrix is $\bSigma = \E [\ybu_m \ybu_m ^H] = \Sbu \bGamma \Sbu^H + \sigma_w ^2 \Ibu$. 
To find $\bgam$ in~\eqref{eq:Y_mat}, 
the maximum likelihood (ML) estimation is then formulated by~\cite{Yu:phase}
\begin{equation}\label{eq:MLE}
	\widehat{\bgam} = \underset{\bgam}{\arg\min} \  \log | \bSigma | + \rm{tr} \left( \bSigma ^{-1} \widehat{\bSigma} \right)
	\ \mbox{subject to }  \bgam \ge 0,
\end{equation}
where $\widehat{\bSigma} = \frac{1}{M} \Ybu \Ybu^H$,
and $\bgam \ge 0$ means that each element of $\bgam$ is non-negative,
due to $\bgam_n =  g_n^2 \abu_n $ for all $n = 1, \cdots, N_d$.
Note that the covariance-based ML estimation problem~\eqref{eq:MLE} 
aims to estimate the device activities and the channel statistics $\bgam$.
With a sufficiently large number of BS antennas,
it is shown in~\cite{Yu:phase} that 
if the number of active devices is $K = \mO(L^2)$,
the reliable performance of
joint activity and data detection can be guaranteed
for randomly generated $\Sbu$
by the solution of the 
ML estimation problem~\eqref{eq:MLE}.

Although the optimization problem~\eqref{eq:MLE} is non-convex,
several algorithms have been explored for solving it iteratively. 
In particular, it turned out that
coordinate descent based algorithms~\cite{Yu:cov, Cui:stat, Yu:phase}
exhibit excellent performance of ML estimation.
In this paper,
we deploy the coordinate descent algorithm in~\cite{Yu:cov}
for ML estimation, which we call CD-ML,
to solve the problem~\eqref{eq:MLE}. 
In CD-ML, active signature sequences are identified by an
estimate of $\bgam$, or $\widehat{\bgam}$, 
through the coordinate selection rule~\cite{Yu:cov}.
From
$\widehat{\bgam} = (\widehat{\gamma}_1, \cdots, \widehat{\gamma}_N)^T$,
we obtain $\widetilde{\bgam}_n = (\widehat{\gamma}_{(n-1)Q+1}, \cdots, \widehat{\gamma}_{nQ})^T$
for $n =1, \cdots, N_d$.
Denoting it by
$\widetilde{\bgam}_n = (\widetilde{\gamma}_n ^{(1)}, \cdots, \widetilde{\gamma}_n ^{(Q)})^T$,
we have $\widetilde{\gamma}_n ^{(q)} = \widehat{\gamma}_{i}$ with
$n = \lfloor \frac{i-1}{Q} \rfloor +1$ and $q = (i-1) \pmod{Q}+1$
for $i = 1, \cdots, N$.
Note that $\widetilde{\gamma}_n ^{(q)} $ is
the estimated channel statistics 
corresponding to the signature $\sbu_n ^{(q)}$ in~\eqref{eq:Sbu_n}.
Then, for each $n$, 
\begin{equation*}
\begin{split}
\xi_n ^{\rm ML} = \max_{q=1, \cdots, Q } \widetilde{\gamma}_{n} ^{(q)} , \qquad
\widehat{q}_n = \underset{ q=1, \cdots, Q}{\arg \max} \ \widetilde{\gamma}_{n} ^{(q)}.
\end{split}
\end{equation*}
Finally, an estimated indicator vector $\widehat{\abu}_n
= (\widehat{a}_n ^{(1)}, \cdots, \widehat{a}_n ^{(Q)})^T$ for device $n$ is
obtained by $ \widehat{a}_n ^{(q)} = 0$ if $q \ne \widehat{q}_n$, and 
\begin{equation}\label{eq:ML_th}
\widehat{a}_n ^{(\widehat{q}_n)} = \left\{ \begin{array}{ll} 1, & \mbox{if } \xi_n ^{\rm ML} \ge \xi_{\sf th} ^{\rm ML} , \\
	0, & \mbox{otherwise,} \end{array} \right. 
\end{equation}
where $\xi_{\sf th} ^{\rm ML}$ is a threshold for device activity.

\section{Sufficient Condition for Maximum Likelihood Estimation}
In this section, we use the coherence of a signature sequence matrix $\Sbu$
to derive a sufficient condition 
to achieve the true solution 
by solving the ML estimation problem~\eqref{eq:MLE}.

\subsection{Asymptotic Performance Analysis for ML Estimation}
Recall $\Sbu \in \C^{L \times N}$ from~\eqref{eq:Y_mat}, where $N = N_d Q$.
Given an arbitrary matrix $\Sbu$, let $\widehat{\Sbu} \in \C^{L^2 \times N}$ be the Khatri-Rao product 
of $\Sbu^*$ and $\Sbu$, defined by
\begin{equation}\label{eq:Khatri}
\widehat{\Sbu} = \left[(\sbu_1 ^{(1)})^* \otimes \sbu_1 ^{(1)} , \cdots,  (\sbu_{N_d} ^{(Q)})^* \otimes \sbu_{N_d} ^{(Q)} \right],
\end{equation}
where `$\otimes$' denotes the Kronecker product.
In what follows, 
Theorem~\ref{th:asym} presents
a necessary and sufficient condition 
for the ML estimation of~\eqref{eq:MLE} to achieve
a true solution of $\bgam$ in~\eqref{eq:Y_mat} 
with an asymptotically large number of BS antennas,
which is a combination of Theorems 2 and 6 in~\cite{Yu:phase}.


\begin{thr}~(\!\!\!\cite{Yu:phase})\label{th:asym}
Let $\bgam^0 = (\gamma_1 ^0, \cdots, \gamma_{N} ^0)^T$ be a true solution of $\bgam$
in~\eqref{eq:Y_mat}, corresponding to true indicator vectors $\abu_1, \cdots, \abu_{N_d}$,
where $\mZ = \{ i \mid \gamma_i ^0 = 0\}$ denotes the index set of zero elements of $\bgam^0$.
Define 
\begin{equation}\label{eq:NC}
\begin{split}
\widetilde{\mN} = \{ \xbu \mid \widehat{\Sbu} \xbu = {\bf 0}  \}, \qquad
\mC = \{ \xbu \mid x_i \ge 0, \ i \in \mZ  \},
\end{split}
\end{equation}
where $\xbu = (x_1, \cdots, x_{N} ) \in \R^{N}$.
As the number of BS antennas increases infinitely, 
the solution of~\eqref{eq:MLE} goes to the true solution, 
i.e., $\widehat{\bgam} \rightarrow \bgam^0$ as $M \rightarrow \infty$, 
if and only if
$\widetilde{\mN} \cap \mC = \{ \bf 0 \}$. 
\end{thr}

\subsection{Coherence-Based Analysis for ML Estimation}
Now, we use the coherence of an arbitrary matrix $\Sbu$
to derive a sufficient condition for the ML estimation~\eqref{eq:MLE} to achieve 
the true solution $\bgam^0$ with
an asymptotically
large number of BS antennas. 
To facilitate our analysis, we make the following assumption
for the null space of $\widehat{\Sbu}$ in~\eqref{eq:Khatri}. 
\begin{itemize}
\item[A1)] The sign of each nonzero element of $\xbu \neq {\bf 0} \in \widetilde{\mN}$ takes on
$+1$ and $-1$ equally with probability $\frac{1}{2}$ for an arbitrary signature sequence matrix $\Sbu$.
\end{itemize}
First of all, we use ${\rm spark} (\widehat{\Sbu})$ to
give a sufficient condition to achieve
the true solution $\bgam^0$ for any set of $K$ active devices,
where ${\rm spark} (\widehat{\Sbu})$ 
is the smallest number of columns of $\widehat{\Sbu}$ that are linearly dependent.

\begin{thr}\label{th:spark}
Let $\widetilde{\mN}$ be defined as in Theorem~\ref{th:asym} 
for a given matrix $\Sbu$, while
$\mZ$ varies depending on device activity, but has a fixed set size.
Let the number of active devices be $K = | \mZ^c |$,
where $\mZ^c = \mX \setminus \mZ $ for
$\mX = \{1, \cdots, N\}$.
If ${\rm spark} (\widehat{\Sbu}) > K + \delta$,
the solution of~\eqref{eq:MLE} goes to the true solution $\bgam^0$ 
asymptotically,
i.e., $\widehat{\bgam} \rightarrow \bgam^0$ as $M \rightarrow \infty$, 
with probability exceeding $1-2^{-\delta}$ 
for any set of $K$ active devices.
\end{thr}

\iproof: See Appendix~\ref{app:spark}.
\qed

Based on Theorem~\ref{th:spark}, we give a sufficient condition for the ML estimation
to achieve the true solution $\bgam^0$ for an arbitrary matrix $\Sbu$,
which is represented by the coherence of $\Sbu$.

\begin{thr}\label{th:coh2}
Recall that $K$ is the number of active devices.
Given an arbitrary matrix $\Sbu$,
the solution of~\eqref{eq:MLE} goes to the true solution $\bgam^0$ 
asymptotically, 
i.e., $\widehat{\bgam} \rightarrow \bgam^0$ as $M \rightarrow \infty$, 
with probability exceeding $1-2^{-\delta}$ for any set of $K$ active devices,
provided that the coherence of $\Sbu$ satisfies
\begin{equation}\label{eq:mu}
	\mu(\Sbu) < \frac{1}{\sqrt{K+\delta-1}}.
\end{equation}
\end{thr}

\iproof: See Appendix~\ref{app:coh2}.
\qed

From
the Welch's lower bound~\cite{Welch:low},
the coherence of 
the matrix $\Sbu \in \C^{L \times N}$ satisfies
$\mu(\Sbu) \ge \sqrt{\frac{N-L}{L(N-1)}}$, 
which yields $\mu(\Sbu) = \Omega \left(\frac{1}{\sqrt{L}} \right)$ 
for large $N$.
Then, a sufficient condition of 
Theorem~\ref{th:coh2} is that 
if $K = \mO(L)$, the ML estimation~\eqref{eq:MLE} can achieve the true solution successfully 
with high probability, using an asymptotically large number of BS antennas.
We remark that this sufficient condition is far from being necessary, 
because the covariance-based ML estimation is known to be able 
to detect up to $K=\mO(L^2)$ number of active devices~\cite{Yu:phase}. 
Nevertheless,
Theorem~\ref{th:coh2} 
is still useful in indicating that 
if the coherence of $\Sbu$ becomes lower, 
the ML estimation can guarantee reliable activity and data detection  
for more active devices. 
Also, the result is useful for an arbitrary matrix $\Sbu$,
particularly if the entries are generated in a deterministic way.

It is implicitly understood in many prior research works that 
the low coherence of a signature sequence or pilot matrix plays an essential role in
guaranteeing reliable detection performance for grant-free access. 
With this awareness, 
several efforts~\cite{Iimori:blin, Rusu:incoh} 
have been made to achieve low coherence for the matrix via optimization algorithms.
In this paper, the sufficient condition of Theorem~\ref{th:coh2} presents a theoretical justification 
for the low coherence of a signature sequence matrix in the ML estimation, 
which contributes to the novelty of this work.

We would like to point out that 
the sufficient conditions of Theorems~\ref{th:spark} and \ref{th:coh2} are for
the ML estimation problem~\eqref{eq:MLE}, 
exploiting the spark and 
coherence of $\widehat{\Sbu}$ defined in Theorem~\ref{th:asym}, respectively.
Clearly,  
the sufficient conditions can be applied to covariance-based algorithms, e.g., CD-ML,
which attempt to solve the ML estimation problem \eqref{eq:MLE}.
However, 
the conditions cannot be applicable to other algorithms, e.g., AMP-based algorithms,
which deal with a different problem setting of estimating jointly the device activities and the channel realizations.

	\begin{rem}\label{rm:N_L2}
		In this paper, 
		we assume $N = N_d Q > L^2$,
		which is suitable for practical mMTC systems.
		Under the assumption,
		the Zadoff-Chu (ZC) sequences~\cite{Chu:ZC} of prime length $L$
		cannot afford to support the non-coherent access scheme of $N > L^2$, 
		since at most $L(L-1)$ signature sequences are available
		from all cyclic shifts of ZC sequences with distinct roots.
		Thus, the ZC sequences, 
		one of the best known deterministic sequences, 
		will not be considered for signatures in this paper.
		For practical mMTC, we are motivated to construct a large number of new deterministic signature sequences 
		of short lengths, i.e., $N>L^2$, 
		for accommodating a massive number of devices with low signaling overhead.
	\end{rem}

\section{Design of Deterministic Non-Orthogonal Signatures}
In this section, we construct four sets of non-orthogonal signature sequences 
in a deterministic fashion, where each set presents
a signature sequence matrix $\Sbu$ with low coherence.

\subsection{General Framework}
Let $\mV = \{\vbu_1, \cdots, \vbu_B\}$ be a set of $B$ unimodular masking sequences\footnote{
In this section, the element index of each masking sequence is $k = 0, \cdots, L-1$
for convenience of analysis. }
of length $L$, where 
$\vbu_b = (v_{b}(0), \cdots, v_{b} (L-1))^T$ for $b = 1, \cdots, B$.
In what follows, we construct a set of non-orthogonal signature sequences 
using the masking sequences in $\mV$.

\begin{const}\label{cst:Fc}
Let $\Fbu_L = \left[\frac{1}{\sqrt{L}} e^{-\frac{j 2 \pi kl}{L}} \right] $ 
be the $L$-point discrete Fourier transform (DFT) matrix, where $ 0 \le k,l \le L-1$.
Using each masking sequence in $\mV$, define an $L \times L$ matrix by
\begin{equation}\label{eq:msk_b}
\Phibu_b = {\rm diag}(\vbu_b) \cdot \Fbu_L, \qquad b = 1, \cdots, B.
\end{equation}
Concatenating $\Phibu_1, \cdots, \Phibu_B$, we obtain  
a matrix $\Phibu \in \C^{L \times N_s}$ by
\begin{equation}\label{eq:Phibu}
\Phibu = [\Phibu_1, \Phibu_2, \cdots, \Phibu_B] = [\phibu_1, \cdots, \phibu_{N_s}],
\end{equation}
where $N_s = BL$, and $\phibu_n \in \C^{L \times 1}$ 
is a sequence of length $L$ for $n=1, \cdots, N_s$.
\end{const}

In Construction~\ref{cst:Fc},
$\Phibu$ supplies total $N_s = BL$ sequences of length $L$, where
a group of $Q$ sequences is uniquely allocated to each device.
Thus, $\Phibu$ can support
at most $\lfloor N_s/Q \rfloor  $ devices
with its column sequences.
In \eqref{eq:Y_mat}, $\Sbu$ is a submatrix of $\Phibu$,
containing the
$\lfloor N_s/Q \rfloor \cdot Q $ columns of $\Phibu$ as its
signature sequences. 
In what follows, 
we show that the coherence of the signature sequence matrix $\Sbu$ from Construction~\ref{cst:Fc}
is bounded by the maximum magnitude of the inverse
Fourier transforms of masking sequence pairs multiplied element-wise. 

\begin{thr}\label{th:coh}
Let $\mW$ be a set containing the sequences of 
$\vbu_i ^* \odot \vbu_j $, where
$\vbu_i, \vbu_j \in \mV$ for $i < j$.
In other words, 
\begin{equation*}\label{eq:mU}
\begin{split}
\mW  = \{ \wbu_1, \cdots, \wbu_D \} 
 = \{\vbu_1 ^* \odot \vbu_2, \cdots, \vbu_{B-1} ^* \odot \vbu_B \},
\end{split}
\end{equation*}
where $D = \frac{B(B-1)}{2}$. 
For $d = 1, \cdots, D$,
let 
$ \widehat{\wbu}_d = 
\frac{1}{\sqrt{L}} \Fbu_L ^* \wbu_d = (\widehat{w}_d (0), \cdots, \widehat{w}_d (L-1))^T$. 
Then, 
the coherence of $\Sbu$ from Construction~\ref{cst:Fc} 
is given by
\begin{equation}\label{eq:S_coh}
\mu (\Sbu) \le   \max_{ \substack{1\leq d \leq D} } \max_{ \substack{0 \leq l \leq L-1} } 
\left| \widehat{w}_d (l) \right|.
\end{equation}
\end{thr}

\noindent \textit{Proof:} 
See Appendix~\ref{app:c}.
\qed

\subsection{Example Constructions}
We employ several known polyphase sequences with low correlation
for masking operation of Construction~\ref{cst:Fc}. Then,
the design framework
produces four sets of deterministic non-orthogonal signature sequences.
For each set, the coherence bound of a signature sequence matrix $\Sbu$
containing the sequences
is derived by leveraging 
the bounds on character sums~\cite{Wang:new},
which is described in Appendix~\ref{app:char}. 
In essence, the example constructions exploit
the low correlation of the polyphase masking sequences, which causes 
the multiplied masking sequence pair to have a bounded maximum magnitude after the inverse DFT. 
Finally, by Theorem~\ref{th:coh}, the resulting deterministic non-orthogonal sequences ensure 
that the matrix $\Sbu$ 	
has theoretically bounded low coherence.

Before describing the details of example constructions,
we introduce some algebraic concepts, which are necessary for 
understanding the polyphase masking sequences. 
Let $\Z_q = \{ 0, 1, \cdots, q-1 \}$, which denotes 
an integer ring of $q$ elements, and 
$\Z_q ^+ = \Z_q \setminus \{0 \}$.
For prime $p$ and a positive integer $m$,
$\F_q = \{0, 1, \alpha, \alpha^2, \cdots, \alpha^{q-2} \}$ is a finite field of $q=p^m$ elements,
where $\alpha$ is the primitive element of $\F_q$,
and 
$\F_q^* = \F_q \setminus \{0 \}$. 
The \textit{trace} function from $\F_{p^m}$ to $\F_p$
is defined by
\[ 
{\rm Tr}(x)=\sum_{i=0} ^{m-1} x^{p^i},   \qquad x \in \F_{p^m}.  
\]
Also, the \textit{logarithm} over $\F_q$ is defined by
\[ \log_\alpha x = \left \{ \begin{array}{ll} t, & \quad \mbox{if } x = \alpha^t, \ 0 \leq t \leq q-2, \\
	0, & \quad \mbox{if } x = 0.  \end{array} \right.
\]
For more details on finite fields and algebraic foundations, 
readers are referred to \cite{Lidl:FF, GolGong:seq}.

We are now ready to present example constructions
using four polyphase masking sequences.

\subsubsection{Signatures from Cubic Masking Sequences}
In~\cite{Alltop:comp}, 
Alltop presented complex-valued \emph{cubic} sequences with low correlation.
Also, it is straightforward to generalize them to 
a cubic sequence family of odd prime length~\cite{Ye:LAZ}.
We use the cubic sequences for masking operation in Construction~\ref{cst:Fc}.
\begin{df}\label{def:cub}
	For odd prime $L$ and integers $\lambda_1, \lambda_2 \in \Z_L$,
	define a masking sequence $\vbu_b = (v_b(0),  \cdots, v_b (L-1))^T$ by
	\begin{equation*}\label{eq:cub_p}
		v_{b} (k) = \exp \left(\frac{j 2 \pi (\lambda_1 k^3 + \lambda_2 k^2) }{L} \right), 
	\end{equation*}
	where $\lambda_1 = \lfloor \frac{b-1}{L}\rfloor $ and $\lambda_2 = (b-1) \pmod{L} +1$.
	We construct a masking sequence set 
	$\mV_C = \{\vbu_1, \cdots, \vbu_{L^2} \}$, 
	where 
	the set size is 
	$B =|\mV_C| = L^2$.
	Using $\mV_C$, 
	Construction~\ref{cst:Fc} gives a set of signature sequences 
	from \eqref{eq:msk_b} and \eqref{eq:Phibu}, 
	where 
	the total number of signatures is 
	$N_s = L^3$.
\end{df}

From the studies of \cite{Alltop:comp} and \cite{Ye:LAZ},
it can be seen that
the cubic based sequences of Definition~\ref{def:cub} are not new, and 
it is easy to show that the corresponding signature sequence matrix 
has theoretically bounded low coherence.

\begin{thr}\label{th:cb_coh}
	Assume that 
	each device has its unique signature sequence set of size $Q$ in
	$\Sbu \subset \Phibu$ from
	Construction~\ref{cst:Fc}, where
	the masking sequence set is $\mV_C$ from
	Definition~\ref{def:cub}. 
	Then, the maximum number of devices to be supported is 
	$\lfloor N_s/Q \rfloor  = \lfloor L^3/Q \rfloor $.
	Also, the coherence of $\Sbu$ is bounded by 
	\begin{equation*}\label{eq:c_coh}
		\mu (\Sbu) \leq 
		\left\{ \begin{array}{ll} \frac{1}{\sqrt{L}}, & \quad \mbox{ if } N_d \le \frac{L^2}{Q} , \\
			\frac{2}{\sqrt{L}}, & \quad \mbox{ otherwise}. \end{array} \right.
	\end{equation*}
\end{thr}

\noindent \textit{Proof:} 
The proof is straightforward from Theorem 5 of~\cite{Ye:LAZ}.
If $N \le L^2$,
the bound is obtained from  
Lemma 3 of~\cite{Ye:LAZ}.
\qed

\subsubsection{Signatures from Power-Residue Masking Sequences}
In~\cite{Sidel:org}, Sidelnikov presented two classes of polyphase sequences, called
\textit{power-residue (PR)} and \emph{Sidelnikov} sequences.
Construction~\ref{cst:Fc} can use each one for
masking sequences.
\begin{df}\label{def:prs}
For odd prime $L$, 
let $\alpha$ be a primitive element in $\F_L$ and $H>2$ be a positive integer that divides $L-1$.
For integers $\lambda_1 \in \Z_L$ and $\lambda_2 \in \Z_H ^+$,
define a masking sequence $\vbu_b = (v_b(0),  \cdots, v_b (L-1))^T$ by
\begin{equation}\label{eq:prs_p}
	v_{b} (k) = \exp \left(\frac{j 2 \pi \lambda_2 \log_\alpha (k+\lambda_1)}{H} \right), 
\end{equation}
where $\lambda_1 = \lfloor \frac{b-1}{H-1}\rfloor $ and $\lambda_2 = (b-1) \pmod{H-1} +1$.
We construct a masking sequence set 
$\mV_P = \{\vbu_1, \cdots, \vbu_{(H-1)L} \}$, 
where 
the set size is 
$B =|\mV_P| = (H-1)L$.
Using $\mV_P$, 
Construction~\ref{cst:Fc} gives a set of signature sequences 
from \eqref{eq:msk_b} and \eqref{eq:Phibu}, 
where 
the total number of signatures is 
$N_s = (H-1)L^2$.
\end{df}

In Definition~\ref{def:prs},
if $H = L-1$, Construction~\ref{cst:Fc} gives $\Phibu$ of size
$N_s =  (L-2)L^2 $ using $\mV_P$.
Consider $\cbu = (c(0), \cdots, c(L-1) )^T$ with $c(k) = \log_\alpha k \pmod{H}$ for $0 \le k \le L-1$, which
is an \textit{$H$-ary power residue (PR)} sequence~\cite{Sidel:org} of length $L$.
Treating the PR sequence as a \emph{seed}, 
each masking sequence $\vbu_b$ of~\eqref{eq:prs_p} is
a modulated version of the seed sequence $\cbu$ 
with $\lambda_1$-shift and $\lambda_2$-multiple,
which has been directly used as pilots for CS based random access~\cite{Yu:cra}.

\begin{thr}\label{th:p_coh}
Assume that 
each device has its unique signature sequence set of size $Q$ in
$\Sbu \subset \Phibu$ from
Construction~\ref{cst:Fc}, where
the masking sequence set is $\mV_P$ from
Definition~\ref{def:prs}. 
Then, the maximum number of devices to be supported is 
$\lfloor N_s/Q \rfloor  = \lfloor (H-1)L^2/Q \rfloor $.
Also, the coherence of $\Sbu$ is bounded by 
\begin{equation*}\label{eq:p_coh}
\mu (\Sbu) \leq 
\left\{ \begin{array}{ll} \frac{\sqrt{L}+1}{L}, & \quad \mbox{ if } N_d \le \frac{(H-1) L}{Q} , \\
	 \frac{2\sqrt{L}+2}{L}, & \quad \mbox{ otherwise}. \end{array} \right.
\end{equation*}
\end{thr}

\noindent \textit{Proof:} See Appendix~\ref{app:P}.
\qed

\subsubsection{Signatures from Sidelnikov Masking Sequences}
\begin{df}\label{def:Sidel}
For prime $p$ and a positive integer $m$,
let $\alpha$ be a primitive element in $\F_{p^m}$ and
$H$ be a positive integer that divides $L = p^m-1$.
For integers $\lambda_1 \in \Z_L$ and
$\lambda_2 \in \Z_H ^+$, define a masking sequence
$\vbu_{b} = (v_{b}(0), \cdots, v_{b} (L-1))^T$ by
\begin{equation}\label{eq:sid_s}
	v_{b}(k) = \exp \left(\frac{j 2 \pi \lambda_2 \log_\alpha (1+ \alpha^{k+\lambda_1})}{H} \right),
\end{equation}
where $\lambda_1 = \lfloor \frac{b-1}{H-1}\rfloor $ and $\lambda_2 = (b-1) \pmod{H-1} +1$.
We construct a masking sequence set 
$\mV_S= \{\vbu_1, \cdots, \vbu_{(H-1)L} \}$, 
where the set size is $B =  |\mV_S| = (H-1)L$.
Using $\mV_S$, 
Construction~\ref{cst:Fc} gives a set of signature sequences 
from \eqref{eq:msk_b} and \eqref{eq:Phibu}, 
where 
the total number of signatures is 
$N_s = (H-1)L^2$.
\end{df}

In Definition~\ref{def:Sidel}, if $H = L$,  
Construction~\ref{cst:Fc} gives $\Phibu$ of size 
$N_s = (L-1)L^2 $ using $\mV_S$.
Consider
$\cbu = (c(0), \cdots, c(L-1) )^T$ with $c(k) = \log_\alpha (1+\alpha^k) \pmod{H}$ for $0 \le k \le L-1$, which
is an \textit{$H$-ary Sidelnikov} sequence~\cite{Sidel:org} of length $L$.
Treating it as a seed,
each masking sequence $\vbu_b$ of~\eqref{eq:sid_s} is a modulated version of the seed sequence $\cbu$
with $\lambda_1$-shift and $\lambda_2$-multiple.

\begin{thr}\label{th:s_coh}
Assume that each device has its unique signature sequence set of size $Q$ in
$\Sbu \subset \Phibu$ from
Construction~\ref{cst:Fc}, where the masking sequence set is $\mV_S$ from
Definition~\ref{def:Sidel}. 
Then, the maximum number of devices to be supported is 
$\lfloor N_s/Q \rfloor  = \lfloor (H-1)L^2/Q \rfloor $.
Also, 
the coherence of $\Sbu$ is bounded by 
\begin{equation*}\label{eq:s_coh}
	\mu (\Sbu) \leq 
	\left\{ \begin{array}{ll} \frac{\sqrt{L+1}+3}{L}, & \quad \mbox{ if } N_d \leq \frac{(H-1)L}{Q}, \\
		\frac{2\sqrt{L+1}+4}{L}, & \quad \mbox{ otherwise} . \end{array} \right.
\end{equation*}
\end{thr}

\noindent \textit{Proof:} See Appendix~\ref{app:S}.
\qed

\subsubsection{Signatures from Trace Masking Sequences}

A general construction of \textit{trace} sequences
was presented in~\cite{Sidel:mutual}. 
For a masking sequence set,
we consider a special case of the trace sequences
from Corollary 7.3 of~\cite{HellKumar:low}.

\begin{df}\label{def:dBCH}
For odd prime $p$ and a positive integer $m$,
let $\alpha$ be a primitive element in $\F_{p^m}$ and $L = p^m-1$.
For an integer $\lambda_1 \in \Z_{L+1} $,
let $ \theta = 0$ if $\lambda_1  = 0$, and $\theta = \alpha^{\lambda_1 -1}$ otherwise.
For $\theta$ and $\lambda_2 \in \Z_{L}$,
define a masking sequence 
$\vbu_b = (v_b(0), \cdots, v_b (L-1))^T$, where
\begin{equation}\label{eq:dBCH2}
v_b(k) = \exp \left(\frac{j 2 \pi {\rm Tr} \left(  \alpha^{k+\lambda_2} + \theta \alpha^{2(k+\lambda_2)}\right)}{p} \right),
\end{equation}
where $\lambda_1 = \lfloor \frac{b-1}{L}\rfloor $ and $\lambda_2 = (b-1) \pmod{L} $.
Then, we construct a masking sequence set $ \mV_T = \{\vbu_1, \cdots, \vbu_{L(L+1)}\}$,
where the set size is $B = L(L+1)$.
Using $\mV_T$, 
Construction~\ref{cst:Fc} gives a set of signature sequences of size $N_s = L^2(L+1)$. 
\end{df}

Consider $\cbu = (c(0), \cdots, c(L-1) )^T$ with $c(k)= {\rm Tr}(\alpha^k)$ for $0 \le k \le L-1$, which
is a \textit{$p$-ary m-sequence}~\cite{GolGong:seq} of length $L$.
Treating it as a seed, each masking sequence $\vbu_b$ of~\eqref{eq:dBCH2} is an added and then modulated version of 
a pair of seed sequences
with shift and decimation.
As a $p$-ary $m$-sequence is generated by a linear feedback shift register (LFSR) over $\F_p$,
the trace masking sequences of $\mV_T$ 
can be generated by a pair of $m$-stage LFSRs,
which allows low-cost implementation.
For more details on the LFSR implementation of a trace function, 
readers are referred to~\cite{GolGong:seq}.

\begin{thr}\label{th:t_coh}
Assume that each device has its unique signature sequence set of size $Q$ in
$\Sbu \subset \Phibu$ from Construction~\ref{cst:Fc}, 
where the masking sequence set is $\mV_T$ from Definition~\ref{def:dBCH}.
Then, the maximum number of devices to be supported is 
$\lfloor N_s/Q \rfloor  = \lfloor L^2(L+1)/Q \rfloor $.
Also,  
the coherence of $\Sbu$ is bounded by 
\begin{equation*}\label{eq:t_coh}
\mu (\Sbu) \leq  
\left\{ \begin{array}{ll} \frac{ \sqrt{L+1}+2}{L}, & \quad \mbox{ if } N_d \le \frac{L^2}{Q} \\
		\frac{2 \sqrt{L+1}+2}{L}, & \quad \mbox{ otherwise } . \end{array} \right.
\end{equation*}
\end{thr}

\noindent \textit{Proof:} See Appendix~\ref{app:T}.
\qed

\begin{table}[!t]
	\fontsize{8}{10pt}\selectfont
	\caption{Examples of Masking Sequence Seeds}
	\centering
	\begin{tabular}{|c|l|}
		\hline
		Masking Sequences   &  Seed    \\
		\hline
		\hline
		Power Residue  &  $ 0, 0, 2, 16, 4, 1, 18, 19, 6, 10, 3, 9,$ \\
		($L=23, H=22$) & $  20, 14, 21, 17, 8, 7, 12, 15, 5, 13, 11. $ \\

		\hline			
		Sidelnikov     &  $ 6, 17, 5, 2, 11, 13, 18, 21, 4, 19, 1, 9, $\\
		($L=24, H=24$) &  $	0, 22, 15, 10, 20, 14, 12, 8, 7, 23, 3, 16. $ \\

		\hline
		Trace     	   &  $2, 4, 2, 0, 1, 4, 4, 3, 4, 0, 2, 3, $\\
		($L=24, p=5$)  &  $3, 1, 3, 0, 4, 1, 1, 2, 1, 0, 3, 2.  $ \\
		\hline		
	\end{tabular}
	\label{tb:comp}
\end{table}

Table~\ref{tb:comp} presents the seed examples 
for PR, Sidelnikov, and trace masking sequences, respectively.
In specific, 
the seeds of PR and Sidelnikov sequences 
are $\log_\alpha k$ and $\log_\alpha (1+\alpha^k)$,
respectively, while 
the seed of trace sequences is
${\rm Tr}(\alpha^k )$, 
where $ 0 \le k \le L-1$.
Then, each masking sequence is generated by 
cyclic shifts, constant multiples, and/or decimation of the seed sequence.
Finally, the proposed non-orthogonal signatures can be generated on-the-fly 
only using the masking seeds, which will be stored in mMTC devices and a BS.

Each example construction of Section IV.B provides
$\mO(L^3)$ non-orthogonal signature sequences
of length $L$, which 
accommodate $\mO(\frac{L^3}{Q})$ devices in the non-coherent access scheme.
Also, the proposed non-orthogonal sequences ensure that the signature sequence matrix $\Sbu$
has the theoretically bounded low coherence of $\mO(\frac{1}{\sqrt{L}})$,
nearly meeting the Welch bound equality.
Thanks to the low coherence,
the sufficient condition of 
Theorem~\ref{th:coh2} suggests that 
the proposed non-orthogonal signatures
can guarantee reliable activity and data detection for $K = \mO(L)$ active devices 
through the ML estimation~\eqref{eq:MLE}.
Moreover, simulation results of Section V reveal that
the ML estimation achieves the reliable performance of joint activity and data detection
even for $K>L$, using a massive number of BS antennas.

\section{Simulation Results}
In this section, we present
simulation results to demonstrate
the performance of proposed non-orthogonal signatures 
for joint activity and data detection
in massive grant-free access.
In simulations, 
each signature  
of length $L$ has the norm of $\sqrt{L}$ and
the noise variance is set as $\sigma_w ^2 = 0.1$.
As in~\cite{Caire:Non} and \cite{Cui:stat}, we assume that the large-scale fading gain is $g_n = 1$ for 
each device $n$, 
with the prior knowledge of its distance 
from the BS.
Under this assumption, we set $\xi_{\sf th}^{\rm ML} =0.25$
in \eqref{eq:ML_th} 
through numerical experiments.

\subsection{Tested Non-Orthogonal Signature Sequences}
The deterministic design of Section IV
does not produce non-orthogonal sequences of an arbitrary length.
Instead, the cubic and the power residue (PR) based sequences from Definitions~\ref{def:cub} and \ref{def:prs}
take the prime length $L$,
whereas the Sidelnikov and the trace based sequences from Definitions~\ref{def:Sidel} and \ref{def:dBCH}
have $L = p^m-1$ for prime $p$.
To maximize the sequence set size, we choose
$H=L-1$ and $H=L$ for PR and Sidelnikov based sequences, respectively. 
In simulations,
the signature sequence length is $L=23$ for cubic and PR based sequences,
while $L=24$ for Sidelnikov and trace based ones.

For performance comparison, 
we consider some known sequences for signatures, 
where the elements are generated in a random fashion.
First, we use random Gaussian sequences, 
where each element is drawn from the i.i.d. complex Gaussian distribution 
with zero mean and unit variance.
Second, we consider the complex-valued MUSA sequences, 
where each element is randomly taken from the 3-level signal constellation, i.e.,
$\frac{\sqrt{3}}{2} [{\pm 1} {\pm j} , \pm 1, \pm j, 0]$, in Fig.~2(b) of~\cite{Yuan:MUSA}.
Finally, we employ random QPSK signature sequences used in~\cite{Larsson:grant},
where each element is randomly taken from $\frac{1}{\sqrt{2}} (\pm 1 \pm j)$.
To obtain each set of random Gaussian, MUSA, and QPSK sequences of length $L=23$,
we construct an $L \times N$ signature sequence matrix with
the lowest coherence through $10$ random trials,
where we observe that the coherence is much higher than those for
the proposed non-orthogonal signatures.

As additional benchmarks, we consider two non-orthogonal sequences of length $L=23$,
exploiting the optimization techniques of C-SIDCO and U-SIDCO in~\cite{Rusu:incoh}. 
The C-SIDCO presents a complex-valued incoherent frame by solving a coherence minimization problem numerically, 
while the U-SIDCO tackles the problem with 
an additional constraint 
that the entries of a frame have equal magnitude.
From each frame, the frame vectors become non-orthogonal signatures. 
As obtained by optimization algorithms, 
the signature sequences 
can 
provide low coherence for the signature sequence matrix.

To validate the assumption A1) of Section III.B numerically for the tested signatures,
we examined the average ratios
of negative to nonzero elements in $\xbu \neq {\bf 0} \in \widetilde{\mN}$,
where $\widetilde{\mN}$ is the null space of $\widehat{\Sbu}$ in~\eqref{eq:Khatri}.
For each matrix $\Sbu$, 
we check the ratios with $1000$ vectors randomly taken from $\widetilde{\mN}$, where
the chosen parameters are 
$(L, N_d, Q) = (23, 200, 4), (47, 1000, 4), (79, 1000, 8)$, respectively\footnote{
	For the signatures from C-SIDCO and U-SIDCO, 
	we checked the average ratios only for
	$(L, N_d, Q) = (23, 200, 4) $, since 
	the optimization algorithms 
	take very long time to generate the signature sequences for large $N_d$.}. 
	From the numerical test, we observe that 
	the average ratios are nearly $0.5$ for the tested signatures,
	from which the assumption A1) turns out to be valid for the coherence-based analysis.

\subsection{Joint Activity and Data Detection}

To evaluate the performance of proposed non-orthogonal signatures,
we use the CD-ML of Section II.B. For comparison purpose,
we also use the MMV-AMP algorithm proposed in~\cite{Ke:mmv-amp}.
The MMV-AMP carries out joint activity and data detection by
estimating the sparse device activities and
the channel realizations jointly, which is described in Appendix~\ref{app:mmv-amp}.
Recall that the sufficient conditions of Theorems~\ref{th:spark} and \ref{th:coh2} 
cannot be applicable to AMP-based algorithms.
In this paper, the MMV-AMP
algorithm 
is only introduced to evaluate
the detection performance of non-orthogonal signatures
in simulations.

In Section II.A, 
recall that 
the indicator vector $\abu_n$ 
has only a single $1$ 
if device $n$ is active, 
and $\abu_n = {\bf0 }$ otherwise.
The detection algorithms of Section II.B ensure that
its estimate $\widehat{\abu}_n$ 
from each algorithm
also has a single $1$ at most.
Given $\abu_n$ and $\widehat{\abu}_n$,
we define an error indicator $e_n$ for $n = 1, \cdots, N_d$, i.e.,
$e_n  = 1$ if $\widehat{\abu}_n $ differs from $\abu_n $ in at least one position,
or $\abu_n \oplus \widehat{\abu}_n \ne {\bf 0}$, 
while $e_n = 0$ if $\widehat{\abu}_n$ is identical to $\abu_n$, or 
$\abu_n \oplus \widehat{\abu}_n = {\bf 0}$. 
Then, an error vector $\ebu = (e_1 , \cdots, e_{N_d} )$ 
can be created for all devices at each access trial.
Note that the error vector $\ebu$
includes miss detection and false alarm errors 
for activity detection.
In addition, it also contains data detection errors 
for active devices.
Finally, we evaluate the detection performance by averaging
the error probability 
$P_e = \frac{\| \ebu \|_1}{ N_d} $.

\begin{figure}[!t]
	\centering
	\includegraphics[width=0.49\textwidth, angle=0]{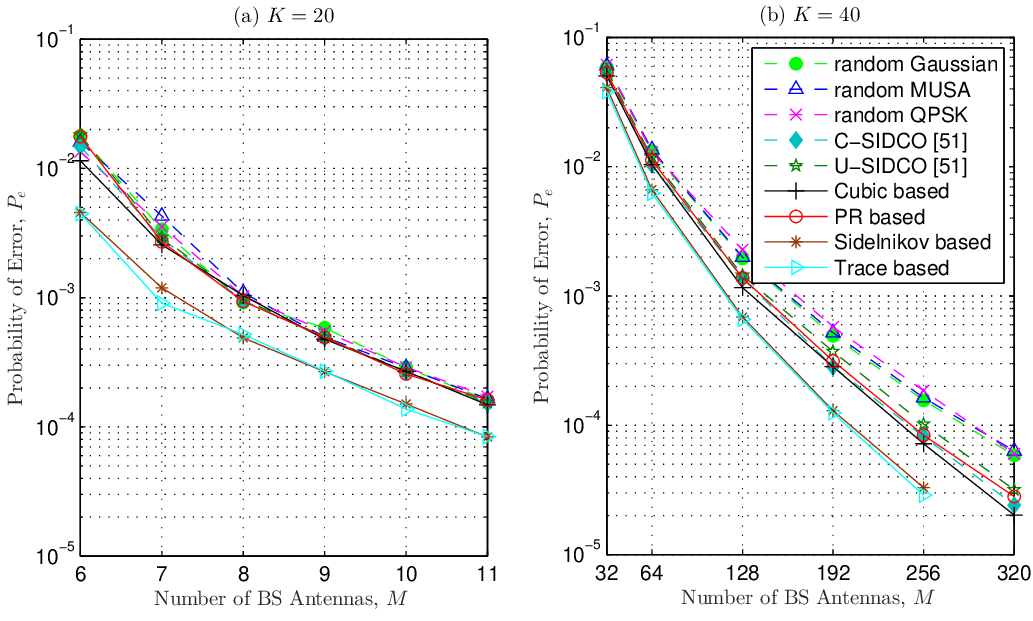}
	\caption{Probability of errors of tested signatures over the number of BS antennas
		by CD-ML, where $N_d = 200, J=2$, and $ Q=4$.
		The number of active devices is (a) $K=20 <L$ and (b) $K=40 > L$, respectively,
		where 
		$L = 23$ for random, C-SIDCO, U-SIDCO, cubic, and PR based sequences, and $L=24$ for Sidelnikov and trace based ones. } 
	\label{fig:ML-L23_K2040}
\end{figure}

Fig.~\ref{fig:ML-L23_K2040} sketches the error probability $P_e$ of
tested non-orthogonal signatures over the number of BS antennas by CD-ML in the non-coherent access scheme,
where $N_d = 200$, $J=2$, and $ Q=4$.  
In the figure, dotted and solid lines correspond to $P_e$
of benchmark and proposed signatures, respectively.
Fig.~\ref{fig:ML-L23_K2040}(a) reveals that 
if the number of active devices is $K=20$, which is less than
the sequence lengths, 
the proposed non-orthogonal signatures have little or no gain over random ones
by CD-ML.
Although the Sidelnikov and the trace based sequences show
less $P_e$ than others, 
it seems to be due to their larger sequence length.
In contrast, Fig.~\ref{fig:ML-L23_K2040}(b) shows that 
if $K=40>L$, the proposed signatures achieve significant performance 
improvements over random ones by
CD-ML using a massive number of BS antennas.
All the proposed signatures 
outperform random ones by CD-ML, thereby
saving at least $40$ BS antennas to
achieve a target error probability, e.g., $P_e=10^{-4}$,
which demonstrates the superiority of proposed non-orthogonal signatures
for joint activity and data detection in massive grant-free access.

\begin{figure}[!t]
	\centering
	\includegraphics[width=0.49\textwidth, angle=0]{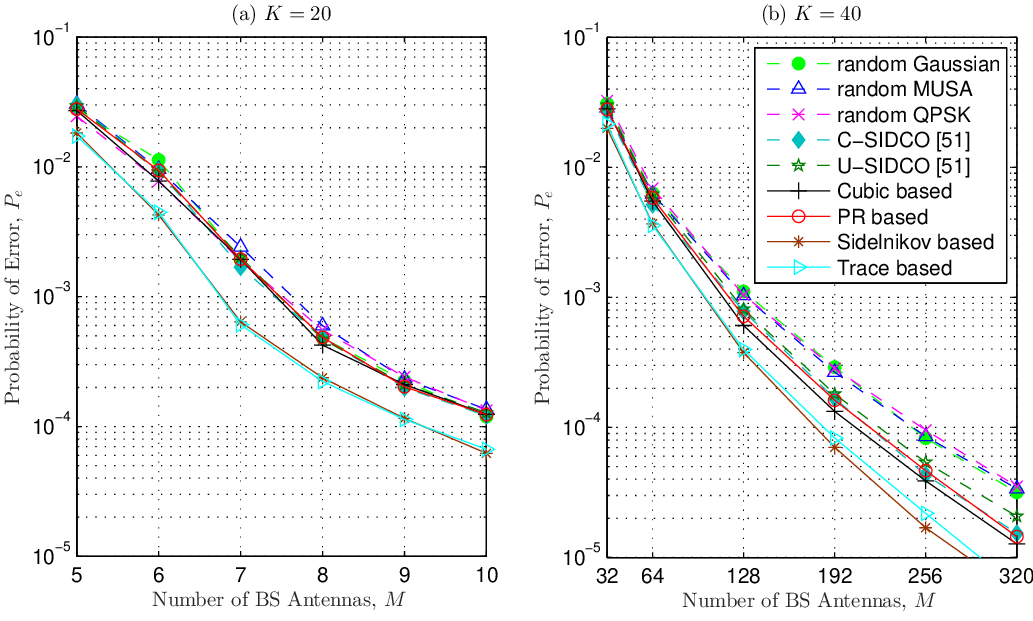}
	\caption{Probability of errors of tested signatures over the number of BS antennas
		by CD-ML, where $N_d = 500, J=1$, and $ Q=2$. 
		The number of active devices is (a) $K=20 < L$ and (b) $K=40 > L$, respectively, where  
		$L = 23$ for random, C-SIDCO, U-SIDCO, cubic, and PR based sequences, and $L=24$ for Sidelnikov and trace based ones.} 
	\label{fig:ML-Nd500_L23_K2040}
\end{figure}

In Fig.~\ref{fig:ML-Nd500_L23_K2040},
we increase the total number of devices to $N_d = 500$
in evaluating the error probability $P_e$ of
tested signature sequences 
by CD-ML,
where $J=1$ and $ Q=2$. 
Similar to Fig.~\ref{fig:ML-L23_K2040}, the proposed
non-orthogonal signatures show no gain over random ones in
Fig.~\ref{fig:ML-Nd500_L23_K2040}(a),
but Fig.~\ref{fig:ML-Nd500_L23_K2040}(b) shows that
for $K>L$,
the proposed signatures outperform random ones apparently via CD-ML.
Thus, the proposed signatures also  
save a number of BS antennas to achieve the target error probability $P_e=10^{-4}$,
when the number of devices is large.

\begin{figure}[!t]
	\centering
	\includegraphics[width=0.49\textwidth, angle=0]{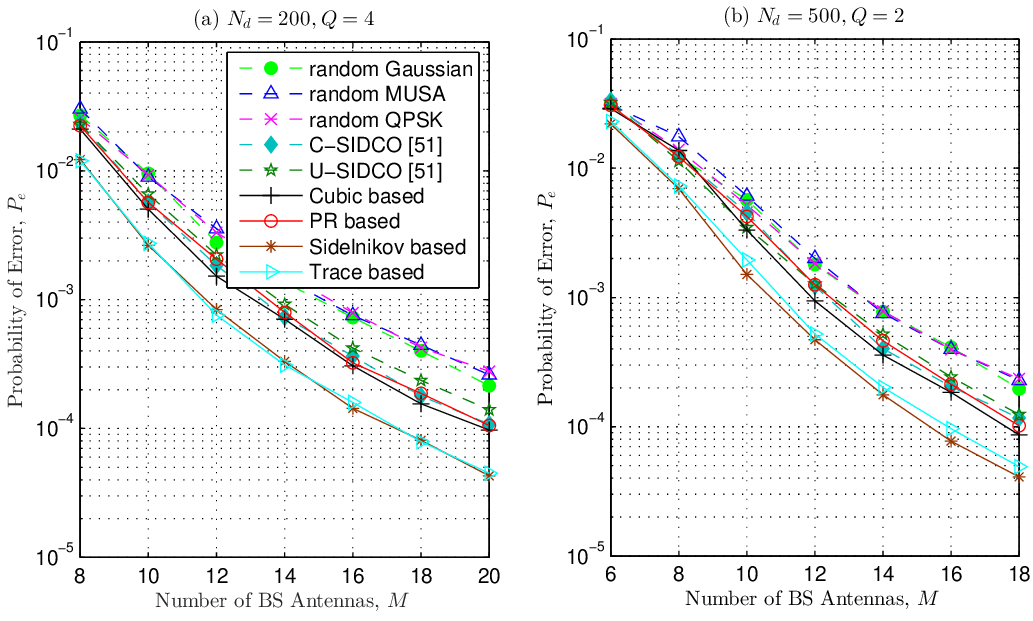}
	\caption{Probability of errors of tested signatures over the number of BS antennas
		by MMV-AMP, 
		where (a) $N_d = 200, J=2, Q=4$ and (b) $N_d = 500, J=1, Q=2$, respectively.
		The number of active devices is $K=20 < L$, where 
		$L = 23$ for random, C-SIDCO, U-SIDCO, cubic, and PR based sequences, and $L=24$ for Sidelnikov and trace based ones.} 
	\label{fig:AMP-Nd_L23_K20}
\end{figure}

Fig.~\ref{fig:AMP-Nd_L23_K20} depicts the error probability $P_e$ of
tested sequences over the number of BS antennas by MMV-AMP, where
(a) $N_d=200$, $J=2, Q=4$ and (b) $N_d = 500, J=1, Q=2$.
In the figure, we set
$K=20 $, 
since the performance of
MMV-AMP is meaningful\footnote{When $K>L$,
we observed that the proposed non-orthogonal signatures beat random ones,
but the error probability of
MMV-AMP flattens out quickly for all the signatures even if $M$ continues to increase.}
for 
$K<L$.
Fig.~\ref{fig:AMP-Nd_L23_K20} shows that 
the proposed signatures outperform random ones by MMV-AMP,
thereby saving a few BS antennas to achieve the target $P_e=10^{-4}$.
In comparison to Figs.~\ref{fig:ML-L23_K2040}(a) and \ref{fig:ML-Nd500_L23_K2040}(a),
we observe that for $K<L$,
the proposed signatures achieve more reliable performance of
joint activity and data detection than random ones 
by MMV-AMP, 
which suggests that the impact of low coherence 
is more outstanding in MMV-AMP than in CD-ML.

\begin{figure}[!t]
	\centering
	\includegraphics[width=0.49\textwidth, angle=0]{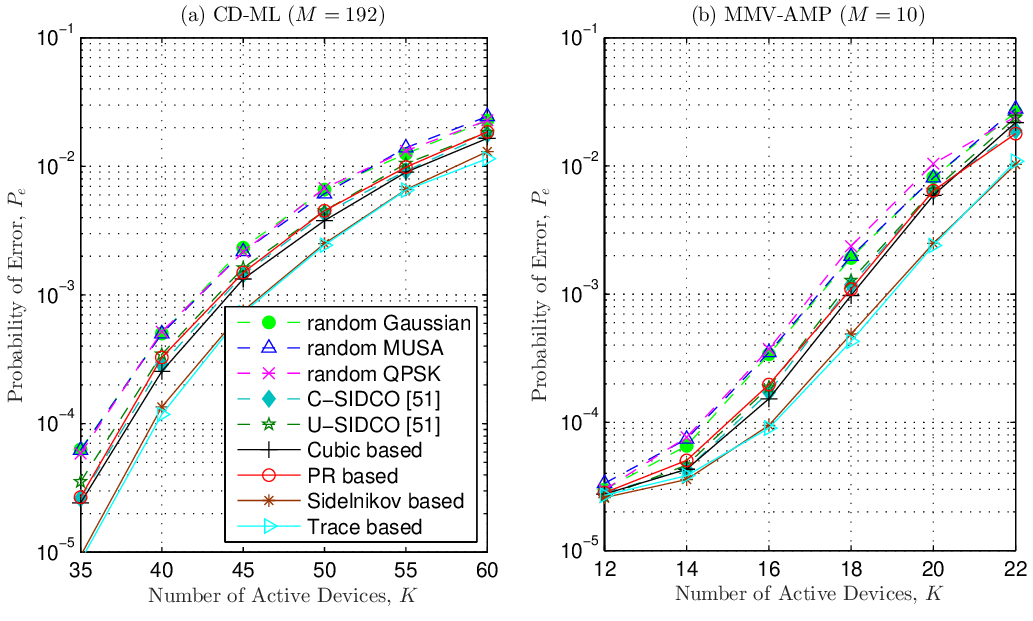}
	\caption{Probability of errors of tested signatures over the number of active devices
		by (a) CD-ML and (b) MMV-AMP,
		where $N_d = 200, J=2$, and $ Q=4$.
		The number of BS antennas is $M=192$ for (a) CD-ML, and $M=10$ for (b) MMV-AMP, respectively. 
		The sequence lengths are
		$L = 23$ for random, C-SIDCO, U-SIDCO, cubic, and PR based sequences, while $L=24$ for Sidelnikov and trace based ones.} 
	\label{fig:K_M192}
\end{figure}

Fig.~\ref{fig:K_M192} sketches the error probability $P_e$ of tested sequences
over the number of active devices, where $N_d=200, J=2$, and $Q=4$.
In Fig.~\ref{fig:K_M192}(a), CD-ML has been deployed for $K>L$ using
a massive number of BS antennas, i.e., $M=192$.
Fig.~\ref{fig:K_M192}(b) shows the performance of MMV-AMP for $K<L$,
where $M=10$. 
Fig.~\ref{fig:K_M192} confirms that the proposed non-orthogonal signatures
are superior to random ones,
achieving less $P_e$ for $K>L$ by CD-ML and for $K<L$ by MMV-AMP, respectively.
In each case, the proposed signatures are able to 
support more active devices than random ones
by achieving the same error probability for massive grant-free access.

Figs.~\ref{fig:ML-L23_K2040}-\ref{fig:K_M192} show
that the performance of the benchmark signatures from C-SIDCO and U-SIDCO
is similar to that of the proposed non-orthogonal signatures.
It is owing to the low coherence of the corresponding signature sequence matrices,
which turned out to be lower than the coherence from the proposed signatures.	
However, the proposed signatures have advantages over the benchmarks
in practical implementation, 
which will be discussed in next subsection.

\subsection{Discussion}

The simulation results of this section
demonstrate that the proposed non-orthogonal signatures
outperform random ones in terms of joint activity and data detection
for massive grant-free access.
The superiority of proposed signatures is evident 
for $K>L$ by CD-ML, and for $K<L$ by MMV-AMP, respectively,
where 
the coherence turns out to be a good indicator for detection performance.
While 
Theorem~\ref{th:coh2}
suggests that the ML estimation~\eqref{eq:MLE} can 
guarantee reliable activity and data detection for $K=\mO(L)$ active devices,
it is noteworthy that the proposed non-orthogonal signatures   
achieve accurate detection even for $K>L$ 
via CD-ML with a massive number of BS antennas.

For joint activity and data detection,
the simulation results confirm that CD-ML always outperforms MMV-AMP in terms of the probability of errors.
Meanwhile, we observe 
that the running time of
CD-ML is much longer than that of MMV-AMP, as witnessed by
Fig.~11 of~\cite{Yu:phase}.
Thus, as long as $K<L$,
MMV-AMP can be a good alternative to CD-ML 
for achieving reliable activity and data detection by consuming
a few more BS antennas, but far 
less computation time.

Further trade-off between detection performance and 
computation time may require the development of
new 
algorithms.
We believe that our deterministic non-orthogonal
signatures will be superior to 
random ones for any type of new algorithms,
as long as the low coherence is exploited effectively.

Further experimental results reveal that the proposed non-orthogonal signatures outperform random ones 
for various lengths, 
but the performance gain 
becomes more
outstanding when the signature length is shorter.
In particular, for given $N$ and $K$, 
the proposed signatures of small $L<K$ clearly outperform random ones
for joint activity and data detection
by CD-ML using a massive number of BS antennas.
As a consequence,
the proposed non-orthogonal signatures of short lengths
will be promising
for massive grant-free access with low signaling overhead.

In practical implementation, 
both mMTC devices and a BS receiver require
storage space for random signature sequences.
As constructed numerically by algorithms, 
the signatures from C-SIDCO and U-SIDCO also
need to be stored at mMTC devices and a BS receiver,
requiring a large amount of storage space, which will be a drawback in practical implementation.
In contrast, the cubic based sequences can be generated on-the-fly.
The PR, Sidelnikov, and trace based sequences
can also be generated on-the-fly, only using their
masking seeds stored in memory. 
For instance, 
random QPSK sequences, which require the smallest storage space
among the benchmarks,
need $2QL$-bit memory for each mMTC device,
while a BS needs to store the sequences of all devices with $2Q L N_d$-bit memory.
Meanwhile, the storage space for a seed of PR or Sidelnikov based sequences
is $L \lceil \log_2 L \rceil$ bits for an mMTC device and a BS, respectively.
Thus, 
if 
$\lceil \log_2L \rceil < 2Q$,
an mMTC device requires less storage space
for the deterministic sequence than for random QPSK.
Moreover, the storage space of a BS is much less than that for random QPSK,
since the signatures of all devices can be generated by the seed.
Consequently, the proposed non-orthogonal signatures
enjoy the benefit of small storage space 
for practical implementation,
thanks to the systematic structure.

\section{Conclusion}

In this paper, we studied the problem of joint activity and
data detection for massive grant-free access in mMTC. We
first derived a sufficient condition for the ML estimation to
solve the problem, which 
is represented by the coherence of the signature sequence
matrix $\Sbu$. 
While it is far from being necessary, the sufficient condition
highlights the importance of low coherence in signature sequence
design. In short, it is desirable to design non-orthogonal signatures such
that the coherence of $\Sbu$ is as low
as possible, which can guarantee reliable activity and data
detection for more active devices via the ML estimation.

Then, we presented a design framework of deterministic non-orthogonal signature sequences. 
Under the framework, we constructed four sets of non-orthogonal sequences.
This deterministic design produces $\mO(L^3)$ sequences of length $L$,
where the coherence of $\Sbu$ 
is theoretically bounded by $\mO(\frac{1}{\sqrt{L}})$.
Simulation results demonstrated that the proposed non-orthogonal
signatures show the outstanding performance of joint activity and data detection
via CD-ML and MMV-AMP, respectively.
In particular, the proposed signatures of short lengths, supported by CD-ML
with a massive number of BS antennas, enable massive grant-free access with low
signaling overhead.
Thanks to the excellent performance and the low implementation cost,
the non-orthogonal signatures based on deterministic design
will be promising for massive grant-free access in mMTC.

%

\appendices

\section{Proof of Theorem~\ref{th:spark}}\label{app:spark}
Let $\mT$ be a support of $\xbu \neq {\bf 0} \in \widetilde{\mN}$.
If $|\mT| = T \le K$, there may exist some $\mZ$ such that
$\mT \subset \mZ^c$.
In this case, 
$\xbu_{\mZ} = {\bf 0}$, leading
to $\xbu  \neq {\bf 0} \in \widetilde{\mN} \cap \mC$
for the corresponding $\mZ$,
which violates the condition of Theorem~\ref{th:asym}.
To avoid this, 
the support set size $T$ should be greater than $K$ 
for all $\xbu \neq {\bf 0} \in \widetilde{\mN}$.

When $T >K$, all elements of $\xbu_{\mZ}$ cannot be zero 
and at least one element of $\xbu_{\mZ}$ must be negative to achieve the condition of Theorem~\ref{th:asym}.
For $\xbu \neq {\bf 0}\in \widetilde{\mN}$,
the number of nonzero elements of $\xbu_{\mZ}$ is $  |\mT \cap \mZ|
= | \mT \setminus (\mT \cap \mZ^c) | \ge T-K $ from $|\mT \cap \mZ^c | \le K$,
where we denote $ T-K \triangleq \delta >0$.
To have $\xbu \neq {\bf 0} \in \widetilde{\mN}$ with its support size $T = K+\delta$,
every subset of $K+\delta$ columns of $\widehat{\Sbu} $ should be linearly independent, 
or equivalently\footnote{While the spark condition is for $\xbu \in \C^{N}$, 
it also holds for $\xbu \in \R^{N}$ in~\eqref{eq:NC}.}
${\rm spark} (\widehat{\Sbu}) > K+\delta$.
Finally, 
if ${\rm spark} (\widehat{\Sbu}) > K+\delta$ ,
at least one element of $\xbu_{\mZ}$ is negative with probability exceeding $1-2^{-\delta}$
under the assumption A1),
which leads to $\widetilde{\mN} \cap \mC = \{ \bf 0 \}$ with high probability for large $\delta$.
\qed

\section{Proof of Theorem~\ref{th:coh2}}\label{app:coh2}
To prove Theorem~\ref{th:coh2}, 
we first study the coherence of $\widehat{\Sbu}$.
\begin{lem}\label{lm:coh}
	Let $\mu(\Sbu)$ be the coherence of an arbitrary matrix $\Sbu$.
	Then, the coherence of $\widehat{\Sbu}$ in~\eqref{eq:Khatri} is given by
	\begin{equation}\label{eq:coh2}
		\mu (\widehat{\Sbu}) = \mu^2(\Sbu).
	\end{equation}
\end{lem}

\iproof:
Let  
$\widehat{\Sbu} = \left[ \widehat{\sbu}_1, \cdots, \widehat{\sbu}_{N} \right]$ from~\eqref{eq:Khatri},
where each column is
\begin{equation}\label{eq:hs}
	\widehat{\sbu}_i = (\sbu_n ^{(q)})^* \otimes \sbu_n ^{(q)}, \qquad  i = 1, \cdots, N
\end{equation}
with $n = \lfloor \frac{i-1}{Q} \rfloor +1$ and $q = (i-1) \pmod{Q} +1$.
In~\eqref{eq:hs}, we denote $i \sim (n, q)$.
Then, the $l_2$-norm of $\widehat{\sbu}_i$ is
\begin{equation}\label{eq:snorm}
	\begin{split}
		\| \widehat{\sbu}_i \|_2 & = \sqrt{\sum_{l=1} ^L \left|s_{n, l} ^{(q)} \right|^2 \cdot \left\| \sbu_n ^{(q)} \right \|_2 ^2 }  
		=  \left \| \sbu_n ^{(q)} \right \|_2 ^2 .
	\end{split}
\end{equation}
Also, the inner product of a pair of distinct columns of $\widehat{\Sbu}$ is
\begin{equation}\label{eq:inn}
	\begin{split}
		\langle \widehat{\sbu}_{i_1}, \widehat{\sbu}_{i_2} \rangle
		& = \sum_{l=1} ^L s_{n_1, l} ^{(q_1)}  \left(s_{n_2, l} ^{(q_2)} \right)^* \cdot
		\left( \sbu_{n_1} ^{(q_1)}\right)^* \sbu_{n_2} ^{(q_2)} \\
		& = \left| \left\langle \sbu_{n_1} ^{(q_1)}, \sbu_{n_2} ^{(q_2)} \right\rangle \right| ^2 ,
	\end{split}
\end{equation}
where $i_1 \sim (n_1, q_1)$ and $i_2 \sim (n_2, q_2)$. 
From~\eqref{eq:snorm} and \eqref{eq:inn},
the coherence of $\widehat{\Sbu}$ is
\begin{equation*}
	\begin{split}
		\mu (\widehat{\Sbu})  & = \max_{ \substack{1 \leq i_1 \neq i_2 \leq N} } \frac{\left| \langle \widehat{\sbu}_{i_1}, \widehat{\sbu}_{i_2} \rangle \right|}
		{\| \widehat{\sbu}_{i_1} \|_2 \| \widehat{\sbu}_{i_2} \|_2} \\
		& = \max_{ \substack{1 \leq i_1 \neq i_2 \leq N} } \left( \frac{\left| \left\langle \sbu_{n_1} ^{(q_1)}, \sbu_{n_2} ^{(q_2)} \right\rangle \right|}
		{\left \| \sbu_{n_1} ^{(q_1)} \right \|_2 \left \| \sbu_{n_2} ^{(q_2)} \right\|_2} \right) ^2 
		= \mu^2(\Sbu),
	\end{split}
\end{equation*}
which completes the proof.
\qed


Based on the result of Lemma~\ref{lm:coh},
Theorem~\ref{th:coh2} can be proven as follows.

\noindent \textit{Proof of Theorem~\ref{th:coh2}}:
By the Gersgorin's circle theorem~\cite{Hogben:hand},
${\rm spark} (\widehat{\Sbu}) > 1 + \mu ^{-1} (\widehat{\Sbu})$~\cite{Eldar:CS}
for an arbitrary $\widehat{\Sbu}$.
Thus, the condition of Theorem~\ref{th:spark} is met
if $1 + \mu ^{-1} (\widehat{\Sbu})
> K+\delta$,
which is equivalent to~\eqref{eq:mu} 
from $\mu (\widehat{\Sbu}) = \mu^2(\Sbu)$. 
\qed

\section{Proof of Theorem~\ref{th:coh}}\label{app:c}
To derive an upper bound on the coherence of $\Sbu$,
we compute the bound of $\Phibu$ instead, 
since $\mu (\Sbu) \le \mu(\Phibu)$ due to $\Sbu \subset \Phibu $.
In~\eqref{eq:Phibu}, we define 
\begin{equation}\label{eq:Gbu0}
\Gbu_{b_1, b_2} = \Phibu_{b_1} ^* \Phibu_{b_2} = 
\Fbu_L^* {\rm diag}(\vbu_{b_1} ^* \odot \vbu_{b_2}) \Fbu_L, 
\end{equation}
where $1\le b_1, b_2 \le B$. 

\noindent \textit{Case 1) $b_1 = b_2 = b$:}
In this case, \eqref{eq:Gbu0} yields  
\begin{equation}\label{eq:G_0}
	\Gbu_{b, b}  
	= \Ibu,
\end{equation}
which clearly shows that the inner product of a pair of distinct columns of $\Phibu_b$ is $\mu_1 = 0$
for any $b = 1, \cdots, B$.

\noindent \textit{Case 2) $b_1 \neq b_2 $:}
In this case, 
each element of $\Gbu_{b_1, b_2} $ in~\eqref{eq:Gbu0} is given by
\begin{equation*}\label{eq:G_3}
	\begin{split}
	\Gbu_{b_1, b_2} (r, c) & = \frac{1}{L} \sum_{k=0}^{L-1} v_{b_1} ^* (k)  v_{b_2} (k) e^{\frac{j2\pi (r-c)k}{L}} \\ 
	& = \frac{1}{L} \sum_{k=0}^{L-1} w_{d} (k)  e^{\frac{j2\pi (r-c)k}{L}} ,
	\end{split}
\end{equation*}
where 
$0 \leq r,c \leq L-1$ and 
$\wbu_{d} = \vbu_{b_1} ^* \odot \vbu_{b_2}$.
From $\widehat{\wbu}_d = \frac{1}{\sqrt{L}} \Fbu_L ^* \wbu_d$, we have
$\widehat{w}_d (l) = \frac{1}{L} \sum_{k=0}^{L-1} w_d ( k)  e^{\frac{j2\pi kl}{L}} $.
Then, it is obvious that 
$ \max_{ b_1, b_2, r,c }| \Gbu_{b_1, b_2} (r,c)|
= \max_{ d,l }  |\widehat{w}_d (l) |$
for $d=1, \cdots, D$ and $l = 0, \cdots, L-1$.
In Case 1), \eqref{eq:G_0} implies 
that each column of $\Phibu$ has unit norm, i.e., $\| \phibu_n \|_2 = 1$
for $n = 1, \cdots, N_s$.
Thus, the coherence of $\Phibu$ in Case 2)
is 
\begin{equation*}
	\mu_2 =\max_{ b_1, b_2, r,c }| \Gbu_{b_1, b_2} (r,c)|
	=  \max_{ 1\leq d \leq D } \max_{ 0 \leq l \leq L-1 } |\widehat{w}_d (l)|.
\end{equation*}
From Cases 1) and 2), 
$ \mu (\Phibu) = \max(\mu_1, \mu_2) = \mu_2$.
Since $\Sbu \subset \Phibu$, we have $\mu(\Sbu) \le \mu (\Phibu) = \mu_2$,
which completes the proof.
\qed

\section{Character Sums and Their Bounds}\label{app:char}

Before presenting the proofs of Theorems~\ref{th:p_coh}-\ref{th:t_coh},
we introduce the basic concepts of characters, character sums, and their bounds,
which will be the background of the proofs.

Let $q=p^m$ for prime $p$ and a positive integer $m$.
For $a \in \F_q ^*$, an \textit{additive} character~\cite{Lidl:FF} of $\F_q$ is defined by
\[
\chi (x) = \exp \left(\frac{j 2 \pi \textrm{Tr}(ax)}{p} \right), \qquad x \in \F_q,
\]
where $\chi(x+y) = \chi (x ) \chi(y)$ for $x, y \in \F_q$.
If $\chi(x) =1$ for all $x \in \F_q$, $\chi$ is a trivial character.

Let $H$ be a positive integer that divides $q-1$.
For $b \in \Z_H ^+$, a \textit{multiplicative} character~\cite{Lidl:FF} of $\F_q $ of order $H$
is defined by $\psi(0) = 0$ and
\[ \psi (x) =  \exp \left(\frac{j 2 \pi b \log_\alpha x}{H} \right) , \qquad x \in \F_q ^*,
\]
where $\psi(xy) = \psi (x) \psi(y)$ for $x, y \in \F_q ^*$.
If $\psi(x) =1$ for all $x \in \F_q ^*$, $\psi$ is a trivial character.
For simplicity, 
we assume $a=1$ and $b=1$ for additive and multiplicative characters, respectively.



Assuming $\psi(0)=1$, 
Facts~\ref{fact5} and \ref{fact4} give useful bounds~\cite{Wang:new} on character sums.
In the following,
$\overline{\F}_q$ denotes the algebraic closure of a finite field $\F_q$ and
$\F_q [x]$ is a polynomial ring over $\F_q$. 
The degree of a polynomial $f(x)$ is denoted as $\textrm {deg}(f)$.

\begin{fact}\label{fact5}\cite{Wang:new}
	Let $\chi$ be a nontrivial additive character of $\F_q$
	and $\psi$ be a nontrivial multiplicative character of
	$\F_q$ of order $H$, respectively, where $\psi(0)=1$.
	For $f(x) \in \F_q[x]$ with $\textrm{deg}(f(x))= r$ and $g(x)\in
	\F_q[x]$, where $g(x)\neq c\cdot h^H(x)$ for
	some $c\in \F_q$ and $h(x)\in \F_q[x]$,
	let $s$ and $e$ be the numbers of distinct roots of $g(x)$ in $\overline{\F}_q$ and $\F_q$, respectively.
	Then,
	\begin{equation}\label{eq:bound4}
		\left|\sum_{x\in \F_q}\chi(f(x))\psi(g(x))\right|\leq
		\begin{array}{ll}
			(r+s-1)\sqrt{q}  +e.
		\end{array}
	\end{equation}
	For the summation over $x \in \F_q^*$, we have 
	\begin{equation}\label{eq:bound5}
		\left|\sum_{x\in \F_q^*}\chi(f(x))\psi(g(x))\right| 
		\leq 	(r+s-1)\sqrt{q}+e+1. 
	\end{equation}
\end{fact}

\begin{fact}\label{fact4}\cite{Wang:new}
	Let $\psi$ be a nontrivial multiplicative character of $\F_q$ of
	order $H$, where $\psi(0)=1$. 
	For $g(x)\in \F_q[x]$, where $g(x)\neq c\cdot h^H(x)$ for
	some $c\in \F_q$ and $h(x)\in \F_q[x]$, let $s$ and $e$ be the numbers of
	distinct roots of $g(x)$ in $\overline{\F}_q$ and $\F_q$, respectively. 
	Then,
	\begin{equation}\label{eq:bound3}
		\left|\sum_{x\in \F_q^*}\psi ( g(x) )\right| \leq 
			(s-1)\sqrt{q}+e+1,   
	\end{equation}
	
\end{fact}
For more details on character sums and their bounds, 
readers are referred to \cite{Lidl:FF, Wang:new}.

We are now ready to present the proofs of Theorems~\ref{th:p_coh}-\ref{th:t_coh}
using Facts~\ref{fact5} and \ref{fact4}.
In Appendices~\ref{app:P}-\ref{app:T},
the proofs are based on the DFT spectra analysis of~\cite{Wang:new}.

\section{Proof of Theorem~\ref{th:p_coh}}\label{app:P}
In Definition~\ref{def:prs}, 
each PR masking sequence is represented by 
$ v_{b}(k) =  \psi \left(g_{\lambda_1, \lambda_2}(x) \right)$,
where $\psi$ is a multiplicative character of order $H$ and
$g_{\lambda_1, \lambda_2} (x) = (x+\lambda_1)^{\lambda_2}$
at $x = k \in \F_L$.
For a pair of distinct masks $\vbu_{b}$ and $\vbu_{b'}$, 
let $ v_{b} (k)  =   \psi \left(g_{\lambda_1, \lambda_2}(x) \right)$ and
$v_{b'} (k)  =   \psi \left(g_{\lambda_1', \lambda_2'}(x) \right)$.   
Then, 
$w_d (k) = 
v_b ^* (k)  v_{b '}(k)$ is represented by
\begin{equation*}
w_d (k)  
= \psi \left({g}_{\lambda_1, H-\lambda_2} (x) g_{\lambda_1', \lambda_2'}(x) \right)
= \psi \left(g(x) \right),
\end{equation*}
where ${g}_{\lambda_1, H-\lambda_2} (x) = (x+\lambda_1)^{H-\lambda_2}$,
$g_{\lambda_1',\lambda_2'} (x) = (x+\lambda_1')^{\lambda_2'}$, 
and $g(x) = g_{\lambda_1, H-\lambda_2}(x)g_{\lambda_1',\lambda_2'} (x) $, respectively.

Since $\textrm{Tr}(x) = x$ for $x \in \F_L$, 
we have
$\exp \left(\frac{j 2 \pi kl}{L} \right) = \chi (lx)$ at $x = k \in \F_L$,
where $\chi$ is an additive character of $\F_L$
and $f(x) = lx$ has the degree of $r = 1$. Thus,
\begin{equation*}\label{eq:p3}
\widehat{w}_{d}(l) 
= \frac{1}{L} \sum_{k=0} ^{L-1} w_{d}(k) e^{\frac{j 2 \pi kl}{L}}
= \frac{1}{L} \sum_{x \in \F_L} \chi (lx)  \psi \left(g(x) \right).
\end{equation*}

If the number of signatures in $\Sbu$ is at most $(H-1)L$, i.e.,
$N = N_dQ \le  (H-1)L$, it means that
$\vbu_b$ and $\vbu_{b'}$ are from a subset of $\mV_P$
containing the first $H-1$ masking sequences, i.e., $b, b' \le H-1$,
which yields
$\lambda_1 = \lambda_1' = 0$. Then,
$g (x)  =  x^{H- \lambda_2+\lambda_2'}$
has $s=1$ and $e=1$ roots in $\overline{\F}_{L}$ and $\F_{L}$, respectively,
where $\lambda_2 \neq \lambda_2 '$ for distinct masks $\vbu_b$ and $\vbu_{b'}$.
Thus, for any $d$ and $l$, 
\eqref{eq:bound4} of Fact~\ref{fact5} yields
\begin{equation}\label{eq:p3_mag1}
\begin{split}
	\left|  \widehat{w}_{d}(l) \right| & =
	\frac{1}{L} \left| \sum_{x \in \F_L} \chi \left( lx \right) 
	\psi \left( g(x) \right) \right| 
	\leq \frac{ \sqrt{L} +1}{L}.
\end{split}
\end{equation}
Meanwhile,
if $ N_d Q >  (H-1)L$,
${g} (x)
= (x+\lambda_1)^{H-\lambda_2} (x+\lambda_1')^{\lambda_2'}$ has $s=2$ and $e=2$ roots 
in $\overline{\F}_L$ and $\F_L$, respectively.
Then, \eqref{eq:bound4} of
Fact~\ref{fact5} gives
\begin{equation}\label{eq:p3_mag2}
	\begin{split}
		\left|  \widehat{w}_{d}(l) \right|
		\leq \frac{ 2\sqrt{L} +2}{L}. \\
	\end{split}
\end{equation}
From \eqref{eq:p3_mag1} and \eqref{eq:p3_mag2},
\eqref{eq:S_coh} yields the coherence bound of $\Sbu$.
\qed

%
\section{Proof of Theorem~\ref{th:s_coh}}\label{app:S}
In Definition~\ref{def:Sidel}, each Sidelnikov sequence is represented
by $v_{b}(k) = \psi \left(g_{\lambda_1, \lambda_2}(x) \right)$,
where $\psi$ is a multiplicative character of order $L$,
$g_{\lambda_1, \lambda_2} (x) = (1+\alpha^{\lambda_1} x)^{\rho \lambda_2}$
at $x = \alpha^k \in \F_{p^m} ^*$, and $\rho = \frac{L}{H}$ is a positive integer.
For a pair of distinct masks $\vbu_b$ and $\vbu_{b'} $, 
let $ v_b (k) =  \psi \left(g_{\lambda_1, \lambda_2}(x) \right)$ and
$ v_{b'} (k) =  \psi \left(g_{\lambda_1', \lambda_2'}(x) \right)$. 
Then, $w_{d}(k) = v_b ^* (k)  v_{b'}(k) =  
\psi \left(\bar{g}_{ \lambda_1, \lambda_2} (x) g_{ \lambda_1', \lambda_2'}(x) \right)$,
where $\bar{g}_{ \lambda_1, \lambda_2} (x) = (1+ \alpha^{\lambda_1} x)^{L- \rho \lambda_2}$ and
$g_{ \lambda_1', \lambda_2'} (x) = (1+ \alpha^{\lambda_1'} x)^{\rho \lambda_2'}$.
Letting $\exp \left(\frac{j 2 \pi kl}{L} \right) 
= \psi (x^{l})$ with $x = \alpha^k \in \F_{p^m} ^*$,
\begin{equation*}
\widehat{w}_d(l) 
= \frac{1}{L} \sum_{k=0} ^{L-1} w_d(k) e^{\frac{j 2 \pi kl}{L}} 
= \frac{1}{L} \sum_{x \in \F_{p^m} ^*}  \psi \left(g(x) \right),
\end{equation*}
where $g(x) = \bar{g}_{ \lambda_1, \lambda_2} (x) g_{ \lambda_1', \lambda_2'} (x) \cdot x^l $.

If $N=N_d Q \le  (H-1)L $, it means that
$\vbu_b$ and $\vbu_{b'}$ are from a subset of $\mV_S$
containing the first $H-1$ masking sequences, i.e., $b, b' \le H-1$, 
which yields $\lambda_1 = \lambda_1' = 0$. 
Then, $g(x) =  (1+  x)^{L- \rho \lambda_2+\rho \lambda_2'} x^l$
has at most $s=2$ and $e=2$ roots in $\overline{\F}_{p^m}$ and $\F_{p^m}$, respectively,
where $\lambda_2 \neq \lambda_2 '$ for distinct masks $\vbu_b$ and $\vbu_{b'}$.
Thus, for any $d$ and $l$, \eqref{eq:bound3} of Fact~\ref{fact4} yields
\begin{equation}\label{eq:s3_mag1}
\begin{split}
	\left|  \widehat{w}_{d}(l) \right|
	\leq \frac{1}{L} \left| \sum_{x \in \F_{p^m} ^*}  \psi \left(g(x) \right) \right|
	\leq \frac{\sqrt{L+1}+3}{L} .
\end{split}
\end{equation}
Meanwhile, if $N_dQ > (H-1)L$, 
$g(x) = (1+ \alpha^{\lambda_1} x)^{L- \rho \lambda_2} (1+ \alpha^{\lambda_1'} x)^{\rho \lambda_2'} x^l$ 
has at most $s=3$ and $e=3$ roots in $\overline{\F}_{p^m}$ and $\F_{p^m}$, respectively.
Therefore, \eqref{eq:bound3} of Fact~\ref{fact4}
yields
\begin{equation}\label{eq:s3_mag2}
\begin{split}
	\left|  \widehat{w}_{d}(l) \right|
	\leq \frac{2\sqrt{L+1}+4}{L} .
\end{split}
\end{equation}
From \eqref{eq:s3_mag1} and (\ref{eq:s3_mag2}), the proof is completed by \eqref{eq:S_coh}. 
\qed

\section{Proof of Theorem~\ref{th:t_coh}}\label{app:T}
In Definition~\ref{def:dBCH},
each trace masking sequence is represented by additive characters in $\F_{p^m} $, i.e.,
$ v_{b} (k) = \chi \left(f_{\lambda_1, \lambda_2}(x) \right)$ at $x = \alpha^k \in \F_{p^m} ^*$,
where $f_{\lambda_1, \lambda_2} (x)= \alpha^{\lambda_2} x + \theta \alpha^{2\lambda_2} x^2 $.
For a pair of distinct masks $\vbu_b$ and $\vbu_{b'} $,
let $v_b(k) =\chi \left(f_{\lambda_1, \lambda_2}(x) \right)$ and  
$v_{b'}(k) =  \chi \left(f_{\lambda_1', \lambda_2'}(x) \right)$.
Then, 
$ w_{d }(k) = v_b ^* (k) v_{b'}(k) =
\chi \left(-f_{\lambda_1, \lambda_2}(x) + f_{\lambda_1', \lambda_2'} (x) \right)$.
With $\exp \left(\frac{j 2 \pi kl}{L} \right) = \psi (x^l)$ at $x = \alpha^k \in \F_{p^m} ^*$,
\begin{equation*}\label{eq:s3}
\begin{split}
	\widehat{w}_{d}(l)  = \frac{1}{L} \sum_{k=0} ^{L-1} w_{d}(k) e^{\frac{j 2 \pi kl}{L}} 
	 = \frac{1}{L} \sum_{x \in \F_{p^m} ^*} \chi \left(f(x) \right) \psi (x^l ), \\
\end{split}
\end{equation*}
where $f(x) =  -f_{\lambda_1, \lambda_2}(x) + f_{\lambda_1', \lambda_2'}(x)$.

If $N=N_d Q \le L^2$, it means $b,b' \le L$, or $\theta = \theta' = 0$ from Definition~\ref{def:dBCH}, where
$f(x) =  (\alpha^{\lambda_2'} - \alpha^{\lambda_2}) x$ has the degree $r=1$
from $\lambda_2 \neq \lambda_2 '$ for distinct masks $\vbu_b$ and $\vbu_{b'}$.
Also, $x^l$ has $s=1$ and $e=1$ roots
in $\overline{\F}_{p^m}$ and $\F_{p^m}$, respectively.
Thus, \eqref{eq:bound5} of Fact~\ref{fact5} yields
\begin{equation}\label{eq:s2}
\begin{split}
	|\widehat{w}_{d}(l)| 
	= \frac{1}{L} \left | \sum_{x \in \F_{p^m} ^*} \chi \left(f(x) \right) \psi (x^l ) \right |
	 \leq 
	\frac{\sqrt{L+1}+2}{L} .
\end{split}
\end{equation}
Meanwhile, if $N_dQ > L^2$,
$f(x) =(\alpha^{\lambda_2'} - \alpha^{\lambda_2}) x  +  (\theta' \alpha^{2\lambda_2 '} - \theta \alpha^{2\lambda_2}) x^2 $ 
has the degree of at most $r=2$,
and $x^l$ has $s=1$ and $e=1$ roots
in $\overline{\F}_{p^m}$ and $\F_{p^m}$, respectively.
Thus, \eqref{eq:bound5} of Fact~\ref{fact5}
gives
\begin{equation}\label{eq:s4}
\begin{split}
	|\widehat{w}_{d}(l) |  \leq 
	\frac{2\sqrt{L+1}+2}{L} .
\end{split}
\end{equation}
Finally, \eqref{eq:s2} and \eqref{eq:s4} complete the proof with \eqref{eq:S_coh}. 
\qed	

\section{Approximate Message Passing (AMP)-Based Estimation}\label{app:mmv-amp}
For joint activity and data detection, 
we describe here a different problem setting of estimating the sparse device activities 
{\it and} the channel realizations jointly using the AMP-based algorithm. 
Note that the covariance-based ML estimation 
of Section II.B
estimates the device activities and the channel statistics only.
	
By defining a channel matrix $\Xbu = \bGamma^{\frac{1}{2}} \Hbu$,
one can translate \eqref{eq:Y_mat} into
\begin{equation}\label{eq:amp_X}
	\Ybu = \Sbu \Xbu + \Wbu.
\end{equation}
Given $\Ybu$ and $\Sbu$,
we 
tackle the \emph{multiple measurement vector (MMV)} problem~\eqref{eq:amp_X} 
to find the row-wise sparse $\Xbu$ for jointly estimating the device activities and the channel realizations.
An important class of algorithms for solving this problem is called
the approximate message passing (AMP)~\cite{Donoho:amp,Ye:BP, Schniter:mmv}, 
which aims to find the minimum mean squared error (MMSE) estimate of $\Xbu$ 
by performing low-complexity message passing
over a bipartite graph.
With a sufficiently large number of BS antennas,
it is shown in~\cite{Liu:mimo,Liu:mimo2} that
the AMP-based algorithm can estimate $\Xbu$ with $K = \mO(L)$ active devices reliably 
for randomly generated $\Sbu$.
Although the theoretical result is derived in an asymptotic regime, 
numerical results reveal that the channel matrix can be estimated accurately 
by the AMP-based algorithm 
for finite $N, K$, and $L$.

In this paper, we use the MMV-AMP algorithm\footnote{We used the code from 
	\url{https://github.com/gaozhen16/Source-Code-M.Ke/blob/main/code_globalsip2018/mmv_amp.m}.}
proposed in~\cite{Ke:mmv-amp}
to estimate $\Xbu=\bGamma^{\frac{1}{2}} \Hbu$. 
While it
resorts to the expectation-maximization (EM) algorithm
to estimate the hyperparameters,
we assume here that the parameters are known 
with the prior knowledge of activity rate, noise variance, and large-scale fading component.
%
When the MMV-AMP returns $\widehat{\Xbu} = [\widehat{\xbu}_1 ^T, \cdots, \widehat{\xbu}_N ^T]^T$
as an estimate of the channel matrix $\Xbu$,
we obtain $\widetilde{\Xbu}_n =$ 
$ [\widehat{\xbu}_{(n-1)Q+1} ^T, \cdots, \widehat{\xbu}_{nQ} ^T ]^T 
\in \C ^{Q \times M}$
for $n=1, \cdots, N_d$.
Denoting it by
$\widetilde{\Xbu}_n=[(\widetilde{\xbu}_n ^{(1)})^T, \cdots, (\widetilde{\xbu}_n ^{(Q)})^T ]^T $,
we have
$\widetilde{\xbu}_n ^{(q)} = \widehat{\xbu}_{i}$ with
$n = \lfloor \frac{i-1}{Q} \rfloor +1$ and $q = (i-1) \pmod{Q}+1$
for $i = 1, \cdots, N$.
Note that $\widetilde{\xbu}_n ^{(q)} $ is
the estimated channel realization across $M$ antennas 
corresponding to the signature $\sbu_n ^{(q)}$ in~\eqref{eq:Sbu_n}.
For each $n$, we compute
\begin{equation*}
\begin{split}
\xi_n ^{\rm AMP} = \max_{q = 1, \cdots, Q} \frac{\| \widetilde{\xbu}_{n} ^{(q)} \|_2 ^2}{M} ,  \qquad
\widehat{q}_n = \underset{ q=1, \cdots, Q}{\arg\max} \ \frac{\| \widetilde{\xbu}_{n} ^{(q)} \|_2 ^2}{M} .
\end{split}
\end{equation*}
Finally, an estimated indicator vector $\widehat{\abu}_n
= (\widehat{a}_n ^{(1)}, \cdots, \widehat{a}_n ^{(Q)})^T$ for device $n$ is
obtained by $ \widehat{a}_n ^{(q)} = 0$ if $q \ne \widehat{q}_n$, and 
\begin{equation*}\label{eq:amp_th}
\widehat{a}_n ^{(\widehat{q}_n)} = \left\{ \begin{array}{ll} 1, & \mbox{if } \xi_n ^{\rm AMP} \ge \xi_{\sf th} ^{\rm AMP} , \\
0, & \mbox{otherwise,} \end{array} \right. 
\end{equation*}
where we set $\xi_{\sf th} ^{\rm AMP} =0.25$ as a threshold for device activity.

\end{document}